\documentclass[12pt, AER]{AEA}

\usepackage[utf8]{inputenc}
\usepackage{amsfonts}
\usepackage{amssymb}
\usepackage{appendix}
\usepackage{enumerate}
\usepackage{euscript}
\usepackage{natbib}
\usepackage{float}
\usepackage{graphicx}
\newtheorem{problem}{Problem}
\newtheorem{proposition}{Proposition}

\setcitestyle{authoryear, open=(,close=)}

\usepackage{xcolor}
\usepackage{amsmath}

\newcommand{\xx}{\boldsymbol{x}}

\newcommand{\zz}{\boldsymbol{z}}

\newtheorem{theorem}{Theorem}
\newcounter{def}
\newtheorem{definition}[def]{Definition}

\DeclareMathOperator{\prlim}{plim}
\newcommand{\plim}{\operatornamewithlimits{\prlim}}

\newcommand{\comm}[1]{}

\newcommand \be  {\begin{equation}}
\newcommand \beno  {\begin{equation*}}
\newcommand \bea {\begin{eqnarray} \nonumber }
\newcommand \ee  {\end{equation}}
\newcommand \eeno  {\end{equation*}}
\newcommand \eea {\end{eqnarray}}
\usepackage{verbatim}
\usepackage{authblk}

\begin{document}

\title{Self-Fulfilling Prophecies, Quasi-Non-Ergodicity \& Wealth Inequality}
\shortTitle{Self-Fulfilling Prophecies}\author{Jean-Philippe Bouchaud and Roger E.A. Farmer\thanks{Bouchaud: Capital Fund Management, Chair of Econophysics \& Complex Systems, Ecole polytechnique, and Acad\'emie des Sciences,  Jean-Philippe.Bouchaud@academie-sciences.fr. Farmer: Department of Economics, University of Warwick and Department of Economics, UCLA, r.farmer.1@warwick.ac.uk. This paper was written  after J. Doyne Farmer suggested that we collaborate as  co-leaders of the Instability Hub for the ESRC funded Network Plus, Rebuilding Macroeconomics.  We thank Angus Armstrong, Pablo Beker, Michael Benzaquen, Leland E. Farmer, Alan Kirman, Robert McKay, Ian Melbourne, Jos\'e Moran, Patrick Pintus and Ole Peters for many insightful discussions on the topic of the paper. The comments of four referees of this journal and the editorial comments of Andrew Atkeson have considerably improved the final draft and we thank all of them for their input.  Thanks also to C. Roxanne Farmer for helpful suggestions and to participants in the University of Virginia macroeconomics seminar in September 2021. We reserve a special thanks to J. Doyne Farmer for his insight that two people with such disparate backgrounds would have something to learn from each other. }}
\date{\today}
\pubMonth{}
\pubYear{}
\pubVolume{}
\pubIssue{}
\begin{abstract}
 We construct a model of an exchange economy in which  agents trade assets contingent on an observable signal, the probability of which depends on public opinion. The agents in our model are replaced occasionally and each person updates beliefs in response to observed outcomes. We show that  the distribution of the observed signal is described by a quasi-non-ergodic process and that people continue to disagree with each other forever. These disagreements generate large wealth inequalities that  arise from the multiplicative nature of wealth dynamics which  make successful bold bets highly profitable. 
\end{abstract}

\maketitle

\begin{quote}
    In standard macroeconomic models rational expectations can emerge in the long run, provided the agents’ environment remains stationary for a {sufficiently long period}. [\citet{EvansHonk2013}].
\end{quote}

\section{Introduction}

Our opening quote from Evans and Honkapohja encapsulates a commonly held view of macroeconomists: that rational expectations is a justifiable assumption because, in a stationary environment,  smart agents are able to learn, after a {\it sufficiently long time}, about the probability distributions of the economic variables they care about.

A stochastic process is a sequence of random variables; it is stationary if the unconditional probability of an element of the sequence is independent of the date at which it is observed and it is ergodic if averages across possible realizations in a given period are equal to the time series average of that variable over many different periods. When ergodicity holds, agents can reliably predict the future by averaging across events that have occurred in the past. Almost all stochastic macroeconomic models are assumed to be ergodic and, for this reason, the argument summarised in our opening quote has proven persuasive to economists who have almost universally adopted the rational expectations assumption since it was introduced into macroeconomics by Robert Lucas fifty years ago \citep{LucasExpectationsNeutrality}.

In the real world, the random events that  influence our lives are neither stationary nor ergodic. But although this observation is banal, it is not  entirely obvious how to construct a model relevant to economics where ergodicity fails. In this paper, we propose a model to explain why agents fail to learn by exploiting the concept of {\em quasi-non-ergodicity} widely used in the physics literature to discuss the properties of glasses and ``spin glasses'' \citep{Anderson_Spin6,Stillinger}.  Quasi-non-ergodicity occurs when a stochastic process is ergodic at very long time horizons,  but where ergodicity breaks down on a time scale at which realizations from the process might realistically be observed by a human agent.\footnote{For physical processes such as glasses and spin glasses, the ergodic time scale can be astronomically long at low temperatures. For the model we construct in this paper, it is longer than the life of most individual human beings.} 
Although the probability distribution of  observable variables is ergodic in the long run; as Keynes famously quipped, ``in the long-run we are all dead''.


To build a quasi-non-ergodic process we assume that agents learn the probability of a bivariate public signal which we refer to as {\em public opinion}. Public opinion is generated as the average probability over the subjective priors of all living agents. Public opinion generates an observable binary random variable that takes one of two values.  By observing a publicly observable sequence of zeros and ones, each individual forms a subjective belief of the time-varying probability that next period's realization will equal $1$. These assumptions lead to a hidden state Markov process in which the state  is a time-varying distribution of subjective probabilities over probabilities.  

To ensure that no agent can exploit ergodicity by being sufficiently patient, we assume that new-born agents do not use the previous history of the public signal. Instead, they begin by making naive forecasts that become increasingly more sophisticated as agents accumulate observations on the signal over time. We show, in this environment, that it is {\em reasonable} for agents to infer the probability of the public signal using a common constant gain learning rule with gain parameter $\lambda$. We elaborate on this idea in Section \ref{Dis} and especially in Section \ref{Reasonable}, where we show that, if all other agents learn with gain parameter $\lambda$,  using the same learning rule as all other agents  generates a forecast that has a negligible bias that is difficult or impossible to detect for almost all values of the time varying probability of the public signal. 

We endow our probabilistic world with a market that allows agents to trade two  securities that are contingent on realizations of the  public signal. At each date, agents solve an inter-temporal optimization program to determine how much of each  security they wish to hold and, because agents have different beliefs, they are willing to trade with each other. 

In the absence of a cost to acquiring information of the kind discussed in \citet{gro-sti-1}, one might expect the market clearing price to reveal the average belief and cause traders to coordinate on the true probability. But in our model, even though there are no costs to information acquisition,  this is not the case. We show that the probability implied by market prices is a {\it wealth-weighted} average of individual subjective beliefs, and not the unweighted average which corresponds, in our model, to the true probability. Some people accidentally benefit from the mismatch between these two probabilities and temporarily earn higher returns from trading in the asset markets. Interestingly, the resulting distribution of wealth is so unequal that the  probability implied by market prices is influenced by the wealthiest agents and fails, even asymptotically, to reveal the true probability of the public signal. Because markets fail to aggregate private information correctly, market prices cannot be used by individuals to reveal the truth.

The wealth distribution generated by our model leads to large and empirically plausible  wealth inequalities even though all agents receive the same non-stochastic endowment in every period. We show that the wealth distribution in our model has a Pareto tail as a consequence of the multiplicative nature of wealth accumulation and interestingly, we are able to reproduce the empirical value of both the exponent of the Pareto tail and of the Gini coefficient of real world wealth distributions using model parameters within a wide interval of reasonable values.\footnote{A Pareto tail refers to the ability of the Pareto distribution to approximate the density of a non-negative random variable for values that are two or more standard deviations above the mean.}  

\section{Literature Review} \label{review}
There is an extensive literature on self-fulfilling prophecies in rational expectations models. Early versions of this literature that rely on dynamic indeterminacy are discussed in Farmer's \citeyearpar{FarmerProphesiesBook1999} textbook and more recent models that display hysteresis and steady-state indeterminacy are reviewed in \citet{far-ind-sch} and explored further in \citet{far-imp-bel}. The literature on self-fulfilling prophecies explains how beliefs drive economic fluctuations, but as with all rational expectations models, eventually everybody agrees with everybody else.  Our current paper, in contrast, explains how a large number of agents interacting in a complete set of financial markets can continue to disagree forever.

\citet{blu-eas-1} discuss two reasons why economists have been attracted to the rational expectations assumption. The first is that rational expectations may be a stable fixed point of an out-of-equilibrium learning mechanism.  This explanation for rational expectations is the one followed by \cite{EvansHonkLeanringBook} that we cited in our opening quote and it is dismissed by Blume and Easley, rightly in our view, because  ``positive results are delicate'' and ``robust results are mostly negative'' \citep[page 929]{blu-eas-1}.

The second route to rational expectations discussed by Blume and Easley is  {\em the market selection hypothesis} introduced by \citet{AlchianInfo} and \citet{FriedmanPositive}. According to this approach, ``those who behave irrationally will be driven out of markets by those who behave {\em as if} they were rational''.\footnote{\citet[page 930]{blu-eas-1}.} Blume and Easley construct an economy populated by infinitely lived agents with dynamically complete markets. They show that if there is a Bayesian learner, with the truth contained in the support of her prior, then all traders who survive will have asymptotically correct beliefs.  

Following \citet{blu-eas-1} an extensive literature builds on their main theme \citep{alv-san-1,cog-sar-1,cog-sar-2,bek-esp-1}. In contrast to this literature, we show that when new agents enter the model, and when the stochastic process they are learning about is quasi-non-ergodic, the economy never converges to a rational expectations equilibrium.\footnote{In a related paper to ours, \cite{jar-bor-1} builds a model with two types of agents with distorted beliefs but his model does not allow beliefs to adapt to changing information.  In \cite{Massari}, an interesting scenario is presented where the market selection mechanism fails in the sense that lucky traders become more wealthy than smart traders (as in our model) but prices still manage to remain efficient. Our model is also related to the ``complex game'' model of \citet{galla-farmer}, in which agents become trapped  in chaotic trajectories that never converge.} In our setting, markets do not favor agents with accurate beliefs and prices fail to reveal the true underlying probabilities.  



We are not the first to explore the topic of non-ergodicity for economics. \citet{brock} have shown that interaction effects can trap the economy in a path-dependent state. \citet{Bouchaud2013} has shown that the Random Field Ising model, which has proven useful in physics to understand interactions between particles, can fruitfully be adapted to understand non-market based interactions between human beings.  And \citet{trapping} have shown that ergodicity breaking occurs in models of habit formation. \citet{horst} reviews the literature on ergodicity and non-ergodicity in economic models.\footnote{\citet{peters} has pointed out that identifying time averages over a single trajectory with ensemble averages can lead to misleading conclusions, and that special care should be devoted to the choice of an appropriate, process dependent, utility function. Our model illustrates a  different facet of non-ergodicity, where agents adapt their beliefs based on an observation window much shorter than the time needed to reach ergodicity.}   In contrast to the literature on non ergodicity, we focus on a case where ergodicity is not strictly broken but where the time scale over which it applies may be  longer than the lifetime of a human agent.\footnote{Our model is a close cousin of Kirman's ant model \citep{kirman}, also known as the Moran model \citep{moran} in the theory of population dynamics, for which results concerning the time taken to converge to the ergodic distribution were recently obtained by \citet{Moran_2020}. Similar situations are encountered in business cycle models with self-reflexive confidence effects \citep{morelli}.}


The closest precursor to our paper is \citet{bek-esp-1}. We modify their environment in two ways. First, the process that generates the states is self-referential and leads to a quasi-non-ergodic process. Second, we modify the environment to allow replacement of agents and we endow new agents with a random prior. Our work is  similar to the discrete time stochastic extensions   by  \citet{far-nou-ven-2} and \citet{Far-ass-perpet} of Blanchard's  \citeyearpar{BlanchardFiniteHorizons} perpetual youth model and the stochastic continuous time version of that model in \citet{Garleanu}. The replacement of agents with new people with random priors is central to our demonstration that beliefs never converge.

Although we use the term `beauty contest', our meaning is distinct from the work of \citet{mor-shi-1} in which a beauty contest is modeled as coordination game in which payoffs are interdependent. A related literature, following \citet{ang-lao-1} and \citet{ben-wan-wen-1} refers to `sentiment' to reflect a similar idea. In contrast to both of these literatures, in our work  individuals alive today try to guess what  individuals who will be born in the future will think an asset will be worth in an environment where there is a no fundamental uncertainty of any kind. Furthermore, in our model, in contrast to these alternative approaches, there exists a set of dynamically complete futures markets. 

An important assumption that drives our results  is  that agents use  constant gain learning to update their beliefs as in the work of  \citet{ben-che-1}, 
\cite{ada-mar-nic-1} and \cite{ada-mar-beu-1}.   Unlike those papers, we study a multi-agent economy and we link the true stochastic process to subjective beliefs through the observation of a public signal which depends on average beliefs.  In our model the event probability  is time dependent, agents continue to disagree with one another forever and the asymptotic wealth distribution is non-trivial and displays a Pareto Tail. The random multiplicative growth mechanism that gives rise to this highly skewed wealth distribution is in the same family of models as those considered in \citet{ bouchaud_mezard,ben-bis-zhu-1} and \citet{ben-bis-1}. 


\section{A Two-Outcome, Self-Referential Model}
\label{sec_dynamics}

We will build up our argument in three stages. In stage one (this section and section \ref{characterize}), we describe a game in which agents form beliefs about a binary outcome and we show that our game leads to a quasi-non-ergodic process for the true belief. In stage two (sections  \ref{sec_hbelief}, \ref{EBU} and \ref{Sim}), we embed our agents in an endowment economy and we allow them to trade Arrow securities contingent on the realization of the binary random variable. In stage three (section \ref{DEq}), we show that the contingent securities market can be replaced by debt and equity and that the equilibria of this more realistic version of our model is the same as the model in which agents trade Arrow securities.  Section \ref{wealth} derives the implications of our model for the wealth distribution.

\subsection{The Beauty Contest Game}

We assume that $N$ agents play a game in which each person must forecast the average belief of the other agents about the outcome of a sequence of binary random events $\{ s_t \in  \mathbf{S} \equiv \{0,1\} \}_{t=1}^\infty$. This is a simple version of a game that Keynes introduced in {\em The General Theory} \citep{KeynesGeneralTheory} to motivate his view that the stock market is driven by what he called `animal spirits'. 

We represent the belief held at date $t-1$ by agent $i$ of the probability that $s_{t} = \{1\}$ as $\mathbb{P}_{i,t}(s=\{1\})$ and we model the self-referential nature of beliefs by assuming that the {\it true probability} of the event, $\mathbb{P}_t(s=\{1\})$, is equal to the average belief,\footnote{More generally, one can consider a model where the true probability is a non-linear, sigmoidal function of the average belief: see Appendix \ref{sigmoid}. Many of the results discussed in the bulk of the paper are actually valid in a more general context, though with interesting twists.}
\begin{equation}\label{EQ1}
 \mathbb{P}_t \equiv \sum_{i=1}^N \frac{\mathbb{P}_{i,t}}{N},
\end{equation}
where throughout the paper, we will drop the argument $s=\{1\}$ after $\mathbb{P}$, unless we explicitly need to distinguish the two outcomes.

One interpretation of our model is that people communicate with others on social networks and each person forms an opinion of what other people think by sampling those within her private network. In the limit, when everyone is connected to everyone else, there is a single value for the beliefs of others which equals the average belief over everyone in the population. We refer to  $\mathbb{P}_t$ as {\em public opinion} and we refer to the event $s_t \in \{0,1\}$  as {\em confidence}.  In the terminology of \citet{cassShellSunspots} confidence is a sunspot.

One possible interpretation of the public signal is the action of an influential journalist who writes an opinion piece in a widely read financial newspaper. That opinion piece can be optimistic -- we interpret optimism as the event  $s_t=1$ -- or pessimistic, we interpret pessimism as the event $s_t=0.$  The probability that the journalist will write an optimistic article is equal to the average degree of optimism in the population as measured by public opinion.

In Section \ref{DEq} we provide an interpretation of our model in which the public signal triggers a common decision on the part of firms to pay dividends in period $t$. In the absence of self-referential effects, the payment or non-payment of a dividend would be irrelevant to the value of the firm. In contrast, in our model the decision to pay a dividend triggers trades between agents in the asset markets. 

\subsection{A Model Where Beliefs are Non-Ergodic}  \label{Dis_0}
In this section we construct a model where people are infinitely lived least-squares learners and we show that in this version of our model public opinion is described by a non-ergodic stochastic process.

We assume that people live forever and although  they initially disagree they are exposed to a common sequence of the realizations of a binary signal.   Each person's prior is an independent random draw from a uniform measure on $[0,1].$ The following equation describes how an individual's belief would evolve if  he were to assume that $\mathbb{P}$ is time invariant.
\begin{align}\label{LS1}
 \mathbb{P}_{i,t+1} & = \mathbb{P}_{i,t} \left( 1 - \frac{1}{t}\right) + \frac{s_t}{t}, \\
 \mathbb{P}_{i,1} & = z_{i,0}, \label{LS2}
\end{align}
where $z_{i,0}$ is an independent draw from a uniform measure on $[0,1].$  Using the definition of $\mathbb{P}_t$ from Eq.\ \eqref{EQ1}, it follows that for large $N$, the evolution of $\mathbb{P}_t$ is given by the equation
\begin{align}\label{LS3}
 \mathbb{P}_{t+1} & = \mathbb{P}_{t} \left( 1 - \frac{1}{t}\right) + \frac{s_t}{t},
\end{align}
where $s_t=1$ with probability $\mathbb{P}_t$ and $0$ otherwise. In this case $\mathbb{P}_t$ converges to a number in $[0,1]$, but that number is different for every realization of $\{\mathbb{P}_t\}_{t=1}^\infty.$ This representation of our model is an economic analogue of the P\'olya urn model, a stochastic process that is well known to be non-ergodic \citep{pemantle}. 

Before providing examples of data generated by equations \eqref{LS1}--\eqref{LS3}, we first provide some definitions and we state a result from stochastic process theory: the {\em mean ergodic theorem}.
\begin{definition}
A {\em stochastic process} is a sequence of random variables $\{x_t\}_{t\in \mathbb{N}}$. A stochastic process is {\em stationary} if the joint probability distribution of \[(x_{t_1},x_{t_2},\ldots,x_{t_k}),\] is the same as the probability distribution of \[(x_{t_1+T},x_{t_2+T},\ldots,x_{t_k+T}),\] for all $t_1,t_2,\ldots,t_k,T \in \mathbb{Z}$.
\end{definition}

Let $F$ be the forward shift operator, let $\omega \in \Omega$ be a draw from a stochastic sequence with $\sigma$-field $\mathcal{F}$, let $\mathcal{L}^p(\mathcal{P})$ be the space of $\mathcal{F}-$measurable functions on $\Omega$ and let $\mathcal{P}$ be a probability measure on $(\Omega,\mathcal{F}).$\footnote{$\Omega$ is the space of sequences which take values in a measurable space $\Xi$, with  $\sigma$-algebra $\mathcal{B}$, and  $\mathcal{F}$ is the product $\sigma$-field. Define a measure $\mathcal{P}$ on $(\Omega,\mathcal{F})$ which describes the evolution of a process $\{x_t\}_{t\in \mathbb{N}}$ over time. 
Let $p \in [1,\infty)$ and define $\mathcal{L}^p(\Omega,\mathcal{F},\mathcal{P})$ as the space of equivalence classes
\[
[X] := \{ Y : X = Y \ \mathcal{P}\text{-almost everywhere}\}
\]
of $\mathcal{F}$-measurable functions $X$ such that
\[
\mathbb{E}(|X^p|)  < \infty.
\]}

 Given these definitions we have the following {\em mean ergodic theorem}. 

\begin{theorem}[Mean Ergodic Theorem]\label{th1}
Let $p\in [1,\infty)$. Then for any $f\in \mathcal{L}^p(\mathcal{P})$, the limit 
\[
\lim_{T\rightarrow \infty} \frac{f(\omega) + f(F\omega) + \ldots + f(F^{T-1}\omega)}{T} = g(\omega)
\]
exists in $\mathcal{L}^p(\mathcal{P})$.  Further, the limit $g(\omega)$ is given by the conditional expectation 
\[
g(\omega) = \mathbb{E}_{\mathcal{P}}(f|\mathcal{I}),
\]
where $\mathcal{I}$ is the invariant  $\sigma$-field defined as  
 \[
 \mathcal{I} = \{A \subseteq \Omega: F A = A \}. 
  \] 
\end{theorem}

A stochastic process that satisfies the assumption of the mean ergodic theorem is said to be {\em ergodic for the mean} and when the conditions of the theorem apply to a stochastic process $\{x_t\}_{t\in \mathbb{N}}$, Theorem \ref{th1} implies that sufficiently long time series averages of $x_t$ will converge to the mean of the  marginal stationary distribution of $\mathcal{P}$ at a point in time. Similar concepts can be used to define ergodicity of higher moments and ergodicity of measures.
\begin{figure}[!hbt] 
\begin{center}
\includegraphics[width = \textwidth]{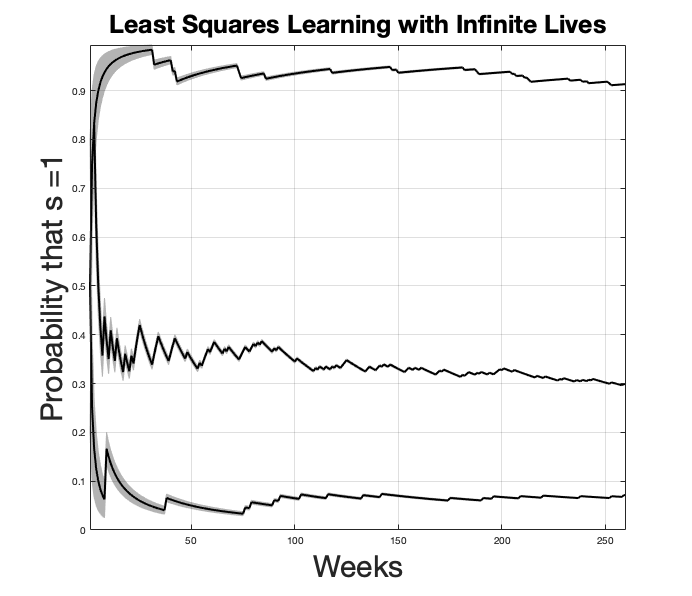}
\caption{The Evolution of Average Beliefs When Agents are Least-Squares Learners with Infinite Lives}\label{Pic1}
\end{center}
\end{figure}

 To illustrate the practical implications of non-ergodicity, Figure \ref{Pic1} plots three different realizations of the process modeled by equations \eqref{LS1} and \eqref{LS2} for an economy with half a million people. The solid black lines are the values of $\mathbb{P}_t$ at each date and the gray shaded areas enclose the  20'th and 80'th percentiles of the distribution of beliefs.  Each run is initialized with 500,000 independent draws from a uniform distribution.

A remarkable feature of these plots is the rapid convergence of opinion;  almost all disagreement vanishes after 75 rounds. But although people converge on a given belief quite rapidly,  they converge to a different value of $\mathbb{P}_t$ for every sequence of draws  $\{s_t\}$.  When people live forever and are least-squares learners, the stochastic process that governs the evolution of $\mathbb{P}_t$ is non ergodic. Although this  example is instructive, it is not very interesting as a theory of why trades take place in asset markets. Everyone's belief eventually converges to the truth and although the truth is itself a function of history, eventually people all agree with one another.

\subsection{A Model Where People Disagree Forever}  \label{Dis}

To generate a theory of permanent disagreement we modify the model in two ways. First, we allow the set of decision makers to change over time by recognizing that people have finite lives. Second, we replace the assumption of least-squares learning with an alternative constant gain learning algorithm in which people discount the far-away past. 

Although one could potentially introduce one of these assumptions without the other, it is natural to make both assumptions together. Suppose, for example, we were to assume that people die but that everyone continues to forecast the future conditional probability using  observations beginning from the date they were born. In this case,  more recent observations will have more weight in the construction of public opinion because  the age distribution of the population will decline exponentially. This observation suggests that a more accurate conditional forecast can be obtained by placing higher weight on more recent observations and that the additional weight placed on more recent observations should be related to the probability of death. Because agents are aware that the world changes, they adapt their learning rule accordingly. 

This intuition is confirmed by numerical simulations. We conducted a series of simulation experiments in which a fraction of the population was assumed to be  constant gain learners and the remaining fraction was assumed to consist of  least-squares learners. We found that for all fractions between 0 and 1, the average mean-square error of the constant-gain learners was substantially lower than that of the least-squares learners. We conclude that agents who place more weight on the recent past will attain  a competitive advantage over least-squares learners and, appealing to that logic, we will assume for the remainder of this paper that agents are  constant gain learners with the same constant-gain learning parameter.\footnote{See section \ref{Reasonable} for a further discussion  of this point.}

To make these ideas precise, we assume that people die with a probability $\delta$ that is independent of age and that when a person dies, she is replaced by a new person with belief $\mathbb{P}_i=z_i$ where $z_i$ is a random variable drawn from a uniform measure on $[0,1]$.  We keep track of who lives and who dies by introducing a random vector $\xx_t \in \mathbf{X} \equiv  \{0,1\}^N$, where  $x_{i,t}=1$ with probability $1-\delta$ and $0$ with probability $\delta$. If a person who was alive in period $t-1$ survives into period $t$ then $x_{i,t}=1$. If she dies then $x_{i,t}=0$. Under these assumptions,  the evolution of the beliefs of the person with index $i$ is given by the expression
\begin{equation}\label{iteration3}
\mathbb{P}_{i,t+1} = x_{i,t}[ (1 - \lambda) \mathbb{P}_{i,t} + \lambda s_{t}] + (1-x_{i,t}) z_{i,t}, 
\end{equation}
where $\lambda \in (0,1)$, $\zz_t \in \mathbf{Z} \equiv [0,1]^N$ and  each element of $\zz_t$ is an independent draw from a uniform distribution.\footnote{The exact form of the distribution of $z_t$ is not important for any of our results. One could in fact assume that $z_{i,t}$ is a weighted sum of the average belief $\mathbb{P}_{t}$ and a uniform random variable. Provided the weight of the latter is non zero, this would not change the structure of the model at all, only the meaning of the parameters. If the weight of the uniform variable is strictly zero, beliefs all converge to either $\mathbb{P}_{i}=0$ or $\mathbb{P}_{i}=1$.}


The term in square brackets on the right side of Eq.\ \eqref{iteration3}  represents the way that a person who is alive in two consecutive periods  updates her belief.  She uses constant gain learning with gain parameter $\lambda$ where a value of  $\lambda$ closer to 1 means that the person puts more weight on recent outcomes. This term is multiplied by $x_{i,t}$ to reflect the fact that it applies only if person $i$ survives into the period. The second term on the right side of Eq.\ \eqref{iteration3} is multiplied by $1-x_{i,t}$.  This reflects the assumption that if agent $i$ dies, her position  is filled by a new-born person who starts life with a random subjective belief, $z_{i,t}$. 

In the limit, as $N \rightarrow \infty$ we can combine  equations \eqref{EQ1} and \eqref{iteration3}  to obtain the following expression for the public opinion,
\begin{equation}\label{EQ2}
\mathbb{P}_{t+1} = (1-\delta)\left[(1 - \lambda) \mathbb{P}_{t} + \lambda s_{t}\right] + \frac{\delta}{2}. 
\end{equation}

\begin{figure}[!hbt] 
\begin{center}
\includegraphics[width = \textwidth]{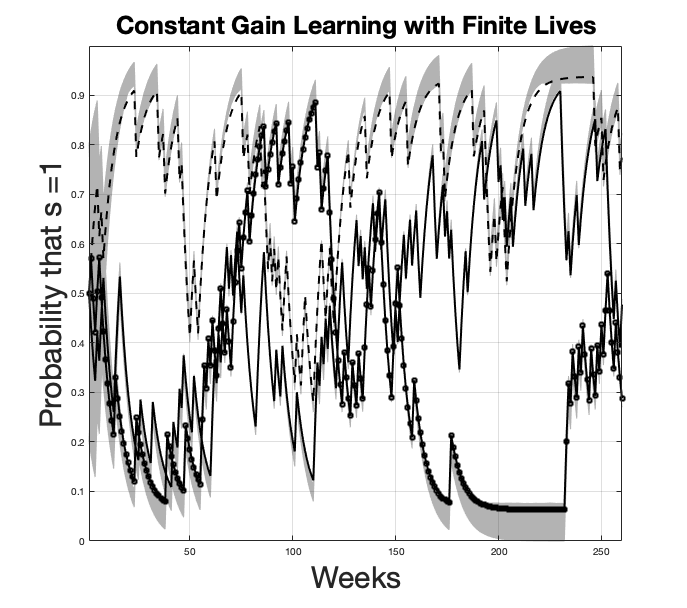}
\caption{The Evolution of Average Beliefs When Agents are Constant-Gain Learners with Finite Lives}\label{Pic3}
\end{center}
\end{figure}

Figure \ref{Pic3} plots three different realizations of the process modeled by equation \eqref{EQ2} for a value of $\delta = 0.02$ and $\lambda = 0.14$. The solid black line, the dashed line and the line marked by circles are the values of $\mathbb{P}_t$ for three different draws from the stochastic process and the gray shaded areas enclose the  20'th and 80'th percentiles of the distributions of beliefs. Our economy  contains 500,000 people and we initialized all three sequences with the same value, $\mathbb{P}_0 = 0.5.$  

These simulations demonstrate that knowledge of public opinion today provides very little information about the state of public opinion in the near future. All three trajectories begin at the same point, but they quickly diverge from each other. 

\subsection{Quasi-Non-Ergodicity}\label{QNE}
In this section we introduce the concept of {\em quasi-non-ergodicity} and we explain how this concept helps us to understand the behavior of the  stochastic sequences depicted in Figure \ref{Pic3}. Our goal is find a way to express the idea that a stochastic process may be ergodic over very long time horizons, but ergodicity may be irrelevant for all practical purposes if the time scale over which convergence is achieved is longer than the lifespan of a human observer. 

To approach this idea we first rewrite the stochastic process defined by equation \eqref{EQ2} as a sequence of probability measures, $\{\mathcal{P}\}_{t=1}^\infty$, generated by the  transition operator $\mathcal{T}$,
\begin{multline}\label{eq_master}
{\mathcal{T}}[\mathcal{P}](\mathbb{P}') \equiv \int_0^1 {\rm d}\mathbb{P} \, \mathcal{P}(\mathbb{P}) \left[\mathbb{P}\, \mathbf{d}\left(\mathbb{P}'-(1-\delta)[(1-\lambda) \mathbb{P} + \lambda] - \frac{\delta}{2}\right)\right.\\ 
 + \left. (1-\mathbb{P}) \, \mathbf{d}\left(\mathbb{P}'-(1-\delta)(1-\lambda) \mathbb{P} - \frac{\delta}{2}\right)
\right].
\end{multline}
Here $\mathbf{d}(\cdot)$ is the Dirac delta function and  the symbols $\mathbb{P}$ and $\mathbb{P}'$ refer to  probabilities in consecutive periods.\footnote{A more usual notation is $\mathbf{\delta}(\cdot)$ for the Dirac delta function. We use $\mathbf{d}(\cdot)$ to avoid confusion with $\delta$, which we reserve for the age-invariant probability of death.}

The measure $\mathcal{P}(\mathbb{P})$ is the probability that $\mathbb{P}\in A$ for any set $A \subset [0,1]$. For given $\mathbb{P}$, $s=1$ with probability $\mathbb{P}$ and $s=0$ with probability $1-\mathbb{P}.$ The Dirac delta function assigns a value to $\mathbb{P}'$  for each of these two outcomes and the measure $\mathcal{P}$ assigns a probability to each of the possible values of $\mathbb{P}$. Integrating over all of these possible values, weighted by $\mathcal{P}$, generates the next period's probability measure $\mathcal{P'}=\mathcal{T}[\mathcal{P}].$

For this definition of $\mathcal{P}$  the probability densities at dates $t$ and $t+1$ are related by the equation
\[
\mathcal{P}_{t+1} = \mathcal{T} \, \mathcal{P}_t ,
\]
and the density at $t$ is related to the initial measure $\mathcal{P}_0$ by the expression,
\[
\mathcal{P}_t= \mathcal{T}^t \,  \mathcal{P}_0,
\]
where $\mathcal{T}^t$ is the $t$'th iterate of the operator $\mathcal{T}$. Notice that $\mathcal{P}$ defines a probability density over probabilities. Complex systems are often defined as probabilistic systems for which probabilities are themselves unknown and must be described with probabilities, as argued by \citet{Parisi2007} and, in an economic context, in \citet{Bouchaud2021}.

Armed with this representation of the system, we are ready to introduce two preliminary concepts, {\em distance from equilibrium} and {\em ergodic time} that we will use to define our central concept: {\em quasi-non-ergodicity}.

\begin{definition} [Distance from Equilibrium]
Let the initial value of a random variable be $x_0 = x_{t=0} \in \Xi$, corresponding to an initial distribution $P_0(x)$ with unit mass localized on $x_0$. The distribution of $x_t$ at time $t$ is obtained from $P_0$ as $P_t = \mathcal{T}^t P_0$. The similarity between the conditional distribution  $P_t$ and the stationary distribution $P_\infty$ can be characterized by a distance $\mathcal{D} \in [0,1]$ defined as \citep{Diaconis}
\[
\mathcal{D}(P_t,P_\infty|x_0) = \sup_{\mathcal{S}} \left| \int_{\mathcal S} {\rm d}P_t -  \int_{\mathcal S} {\rm d}P_\infty \right|,
\]
where $\mathcal S \in \mathcal{B}$ is any subset of $\Xi$. 
\end{definition}
The argument of the sup operator  measures the difference in mass that  $P_t$ and $P_\infty$ attribute to any subset of the space $\Xi$ and the distance between the two measures, represented by $\mathcal{D}$, is small when $P_t$ and $P_\infty$ assign similar probabilities to all possible subsets  of $\Xi$.

Next, we need a way to measure how long it takes for a sequence of probability distributions to converge to an invariant measure. That requirement leads us to define the concept of {\em ergodic time}.
\begin{definition}[Ergodic Time]
An ergodic stationary stochastic process $\{x_t\}_{t \in \mathbb{N}}$ has  ergodic time $T_{\text{e}}(\epsilon)$ if
\[
\int {\rm d}P_\infty(x_0) \, \mathcal{D}(P_{t > T},P_\infty|x_0)) \leq \epsilon.
\]
\end{definition}
In words, for a given confidence level $\epsilon$, the ergodic time $T_{\text{e}}(\epsilon)$ is the time  beyond which the difference between the conditional distribution and the  stationary distribution of the random variable $x$ is no greater than $\epsilon$. Finally, we are ready to introduce our central concept.
\begin{definition}[Quasi-non-ergodicity]
If  $T_{\text{e}}(\epsilon)$ is larger than some large time $K$  we say that the stochastic process $\{x_t\}_{t\in \mathbb{N}}$ is {\em $K-\epsilon-$quasi-non-ergodic}. 
\end{definition}
In the rest of the paper we drop the terms $K$ and $\epsilon$ and refer simply to quasi-non-ergodicity. Informally   $\epsilon$ is a small number and $K$ is greater than the lifespan of a typical observer.

\section{Characterizing the Invariant Measure}\label{characterize}
In this section we characterize the properties of the invariant measure as $N\rightarrow \infty$ and the period length $\Delta t \rightarrow 0.$ We refer to this as the large $N$ continuous time limit. 

Introducing the change of variable  $u = \mathbb{P}-\frac12$, we show in Appendix \ref{newapp} that in the large $N$ continuous time limit, $\mathcal{P}_t(u)$ converges to a symmetric beta-distribution with parameter $\alpha=\delta/\lambda^2$,
\begin{equation}\label{eq_Pbeta}
\mathcal{P}_\infty(u) = \frac{\Gamma(2\alpha)}{\Gamma^2(\alpha)}\left(\frac14 - u^2\right)^{\alpha-1}.
\end{equation}
This distribution is hump-shaped for $\alpha>1$ -- this is the case where $\delta>\lambda^2$ -- and U-shaped when $\alpha<1$ -- this is the case where $\delta<\lambda^2$. In our baseline calibration we choose $\alpha=1$, which coincides with  $\delta=\lambda^2$, but all of our results are robust to variations in $\alpha$ in a wide range of values between $\alpha = 0.5$, for which $\mathcal{P}_\infty(u) $ is U-shaped and $\alpha=2$, for which it is hump-shaped.\footnote{When $\delta \to 0$, the distribution of $\mathbb{P}$ becomes highly peaked around $0$ and $1$. In fact, such long polarisation periods was Kirman's motivation for introducing his ant recruitment model for opinion dynamics \citep{kirman}.  See also \citet{P_Young}.}

The properties of this invariant measure depend on two parameters, $\delta$ and $\lambda$. The parameter $\delta$ is closely related to the ergodic time and  \citet{Moran_2020} have shown that, if we fix $\alpha$ and take $\delta \to 0$ that $T_e(\epsilon)$ is of order $\delta^{-1}$.  The parameter $\lambda$ has a similar interpretation in terms of the {\em memory time}.

\begin{definition}[Memory Time]
Let  person $i$ use the rule 
\begin{equation}\label{rule}
\mathbb{P}_{i,t+1} =  (1 - \lambda) \mathbb{P}_{i,t} + \lambda s_{t}
\end{equation}
to forecast future value of $\mathbb{P}_{T}$ for $T>t$.  The memory time, $T_{\text{m}}(\epsilon)$ is the number of periods after which the  observation $s_t$ has weight less than or equal to $\epsilon$.  It is defined by the expression
\[
(1-\lambda)^{T_{\text{m}}} = \epsilon.
\]
\end{definition}
It follows from this definition that, holding  $\epsilon$ fixed, $T_{\text{m}}(\epsilon)$ is of order $\lambda^{-1}$.

Define  the $i$'th person's {\em degree of disagreement}, $\mathbb{D}_{i,t}$ as the   difference between the belief of agent $i$ and the average belief across all members of the population. In symbols, 
\begin{equation} \label{eq_Ddef}
    \mathbb{D}_{i,t} \equiv \mathbb{P}_{i,t} - \mathbb{P}_t.
\end{equation}
When we fix $\alpha$ and take $\lambda \rightarrow 0$ we are able to obtain an exact expression for the stochastic evolution of  $\mathbb{D}_{i,t}$, in the continuous time limit,
\begin{align}\label{eq_D}
\mathbb{D}_{i,t+1} &= x_{i,t}\left[ (1 - \lambda) \mathbb{D}_{i,t} + \delta \left((1-\lambda) \mathbb{P}_t + \lambda (s_{t}-\frac12) \right)  \right] \\ \nonumber &+ (1-x_{i,t})\left[z_i - (1-\delta)\left((1-\lambda) \mathbb{P}_t + \lambda s_{t} + \frac{\delta}{2}\right) \right].
\end{align}
We show in Appendix \ref{app_dispersion}, that the unconditional expectation of  $\mathbb{D}_i$, converges to zero almost surely  and that, in the large-$N$ --  small-$\lambda$ limit, its variance is given by the expression,
\be\label{eq_dispersion}
\mathbb{V}[\mathbb{D}_i] =  \left[\frac{\lambda}{2 +  (\alpha-1)\lambda} \right]\frac{\alpha(\alpha +2)}{6(2\alpha +1)} \, + O(\lambda^3),  \qquad \alpha= \frac{\delta}{\lambda^2}.
\ee 
The variance of $\mathbb{D}_i$ is a measure of disagreement between agents in the unconditional limiting distribution. For fixed $\lambda$, the disagreement tends to zero as $\alpha \to 0$. In this case $\mathcal{P}_\infty(u)$ has mass points at zero and 1 and  everybody agrees in the limiting distribution that either $\mathbb{P}=0$ or $\mathbb{P}=1$. If $\alpha$ is small, but not equal to zero, public opinion switches between these two mass points with a frequency that vanishes asymptotically as $\alpha \to 0$.

In the opposite case as $\alpha \to \infty$, $\mathcal{P}_\infty(u)$ has a single mass point at  $u=1/2$ and $\mathbb{V}[\mathbb{D}_i]$ converges to the variance of the distribution of initial beliefs. When this distribution is uniform, as we assumed to derive Eq.\ \eqref{eq_dispersion}, $\mathbb{V}[\mathbb{D}_i] \to \frac{1}{12}$. In this case agents never agree because they die  much faster than they  learn. 

\begin{figure}[!hbt] 
\begin{center}
\includegraphics[width = \textwidth]{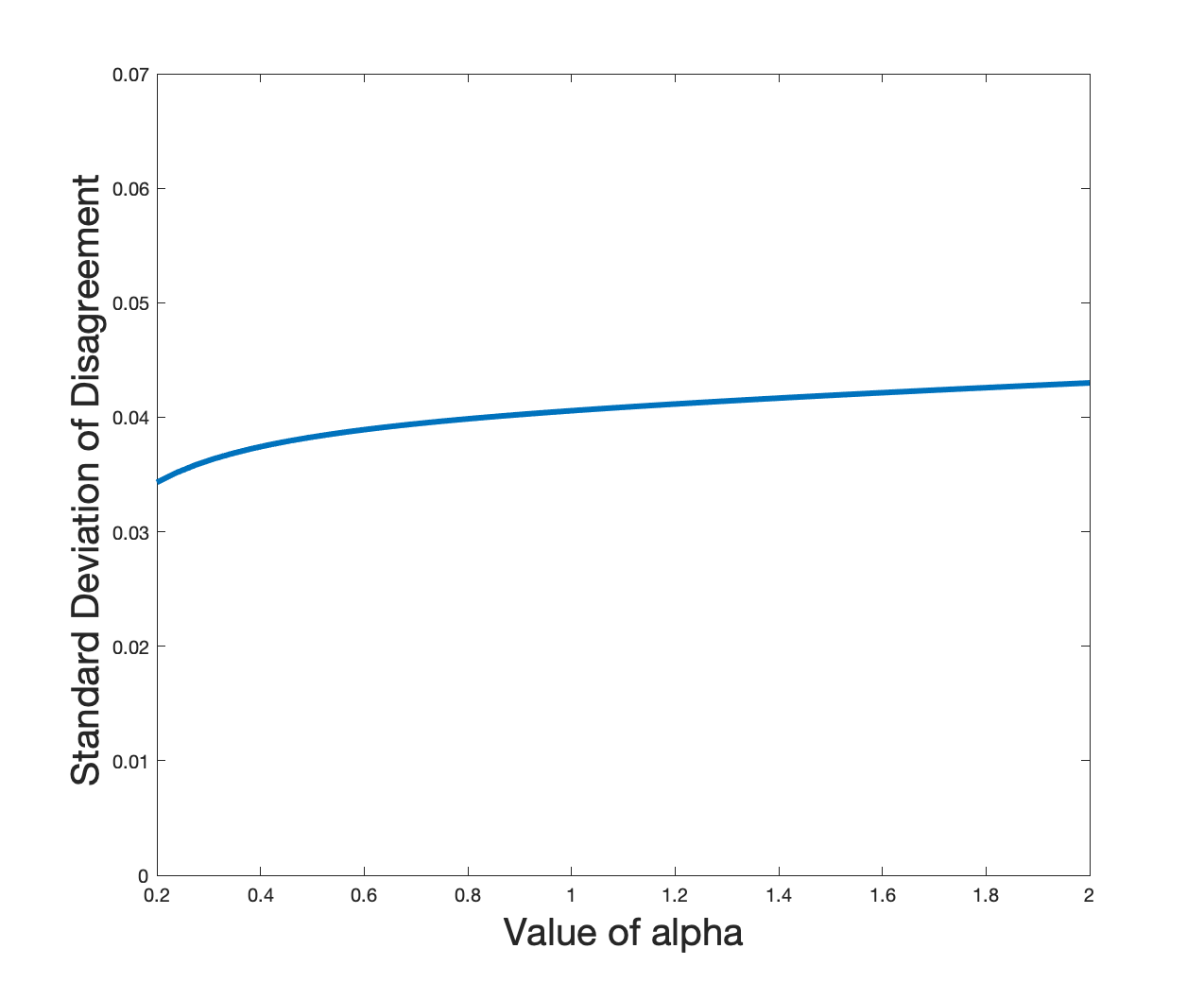}
\caption{How Disagreement Varies with $\alpha$}\label{Fig3}
\end{center}
\end{figure}

In our simulations we chose a time interval of one week and we set  $\delta=3.9 \times 10^{-4}$. These choices imply that  life expectancy, averaged over people of all ages, is approximately $50$ years which accords well with crude estimates from US actuarial tables. For this fixed value of $\delta$, the standard deviation of disagreements is plotted as a function of $\alpha$ in Figure \ref{Fig3}, for $\alpha \in (0.2,2)$. When $\alpha=1$,  the standard deviation of $\mathbb{D}_i$ is approximately  $4 \%$,  corresponding to a memory time $\lambda^{-1}$ of $50$ weeks. 

Figure \ref{Fig3} demonstrates that although the standard deviation of $\mathbb{D}$ varies  with $\alpha$,  it remains within a very small range of approximately $3.5\%$ to $4.5\%$. This level of disagreement is the same order of magnitude as that reported by \citet{Kahneman2021} for the dispersion of the estimates of experts using common information.  As we will show in our simulations, it is large enough to generate substantial discrepancies between the market price and the true price, and a ``fat'' power-law right tail of the wealth distribution when people make bets based on their subjective beliefs.

\subsection{Can Some Agents Learn Better Than Others?}\label{Reasonable}

In this section we explore the question: Is it reasonable to use constant gain learning with gain parameter $\lambda$, given that everyone else in the economy is using the same forecast mechanism? We use the word {\em reasonable} because the use of constant gain learning in this environment is clearly not optimal since the individual learning rule, given by Equation \eqref{iteration3},
\begin{equation}\tag{\ref{iteration3}}
\mathbb{P}_{i,t+1} =  (1 - \lambda) \mathbb{P}_{i,t} + \lambda s_{t}, 
\end{equation}
is different from the evolution of the true probability, given by Equation \eqref{EQ2},
\begin{equation} \tag{\ref{EQ2}}
\mathbb{P}_{t+1} = (1-\delta)\left[(1 - \lambda) \mathbb{P}_{t} + \lambda s_{t}\right] + \frac{\delta}{2}.
\end{equation}
The difference between the individual learning rule and the true evolution of public opinion arises because agents using naive constant gain learning neglect to account for  the  arrival of new agents at rate  $\delta$. 

Consider the problem of a single individual, born at date $j$, who observes the sequence $\{s_t\}_{t=j}^T$. The optimal Bayesian forecast of $\mathbb{P}_t$ is the solution to a non-linear filtering problem where $s_t$ provides a noisy signal of the hidden state variable $\mathbb{P}_t$.\footnote{Although this problem is superficially similar to the problem of forecasting the state in linear state space model, it is complicated by the facts that the variance of  $\{\mathbb{P}_t\}$ is time varying and that the shocks to the state equation and the measurement equation are correlated. The optimal Bayesian forecast could be found using non-linear methods such as the particle filter, but applying methods of this kind are costly.} It follows that constant gain learning is not the best that an arbitrary observer with limitless computational power could achieve. But in the real world people do not have limitless computational power and it may be sufficient to use a simpler rule that has a low predictive error. 

So how bad is the constant gain learning rule? Consider the following representation of this rule which we refer to as the R-estimator,
\[
R_{t+1} = \sum_{j=0}^{t} (1-\lambda)^{t-j}\lambda s_j + (1-\lambda)^{t+1}R_0.
\]
 The following discussion is based on an approximation that is valid for time periods in the interval ${\lambda}^{-1} \ll t \ll {\delta}^{-1}$ which, for the case $\alpha=1$, is between one year and fifty years. On time scales, where $ \lambda^{-1} \ll t$ there is enough data to form estimates of $\mathbb{P}_t$ and for time scales where $t \ll \delta^{-1}$, $\mathbb{P}_t$ is approximately constant. 
 
On these time scales the R-estimator  is conditionally biased, with a bias given by the expression,
\[ 
\mathbb{E}[R_t - \mathbb{P}_t|\mathbb{P}_0] \approx \frac{\delta}{\lambda} \left(\mathbb{P}_0 - \frac12 \right), \qquad \mathbb{P}_0 := \mathbb{P}_{t=0},
\]
where the expectation is taken with respect to the true conditional time-varying probability that $s_t=1$. This bias term reflects the fact that an agent who uses constant gain learning neglects the mean-reverting force towards $\mathbb{P} = 1/2$ which is induced by the birth of new agents with priors centered on $\mathbb{P}_{i,t} = 1/2$.\footnote{Note that  the unconditional mean of $\mathbb{P}_0$ is equal to $1/2$, which implies that $\mathbb{E}[R_T-\mathbb{P}_T | \mathbb{P}_0 \rightarrow 0$ as $T\rightarrow \infty.$ In words, the $T$-step ahead R-estimator is asymptotically unbiased.} 

Could this bias be detected by the agent over time periods of order $\lambda^{-1}$; that is, over lengths of time consistent with the learning window? For that to happen, the agent would need to observe a series of binary outcomes that were statistically implausible given her current belief about the value of $\mathbb{P}_t$. But for most of the range of $\mathbb{P}_t$, the variance of the R-estimator is large relative to its bias. The standard deviation of the R-estimator is approximated by the expression,
\[ 
\text{SD}[R_t] \approx \sqrt{\frac{\lambda}{2} \mathbb{P}_0 \left(1-\mathbb{P}_0\right)}.
\]
To detect the bias, the standard deviation of the estimator must be small relative to the bias. This condition is represented by the inequality,
\[
\sqrt{\frac{\lambda}{2} \mathbb{P}_0 \left(1-\mathbb{P}_0\right)} \ll 
\lambda \left|\mathbb{P}_0 - \frac12 \right|,
\]
where we have used the fact that $\delta = \alpha \lambda^2$, and the special case $\alpha=1$.
\begin{figure}[!thb] 
\begin{center}
\includegraphics[width = \textwidth]{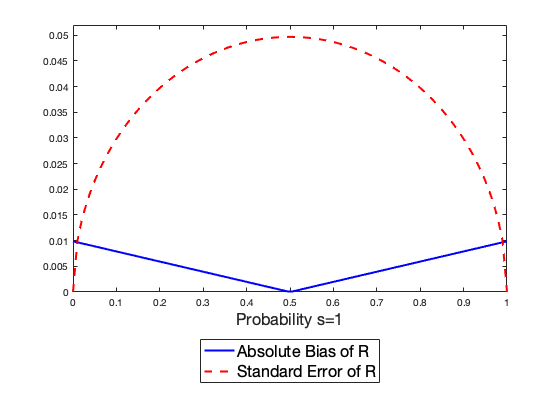}
\caption{A Comparison of the absolute bias of the R-estimator with its standard error }\label{fig0}
\end{center}
\end{figure}
These two terms have a component that depends on $\mathbb{P}_t$ and a component that depends on $\lambda$. In Figure \ref{fig0} we set $\delta = 3.9 \times 10^{-4}$ and  we plot the absolute bias of the R-estimator as the solid line and its standard deviation as the dashed line. Both plots are functions of $\mathbb{P}_t$. This figure makes clear that, except for a sliver of values close to either of the extreme possible values of $\mathbb{P}_t$, the bias of $R_t$ is swamped by its standard deviation.\footnote{We have also constructed this figure for values of $\alpha=0.5$ and $\alpha=2$. The results are qualitatively identical to those we report in Figure \ref{fig0}, reflecting the fact that our results are not sensitive to the value of $\alpha$ for a large range of values that includes our chosen parameterization of $\alpha=1$.}

This result holds because, most of the time, the assumption that $\mathbb{P}_t$ is a random walk is a  good approximation to the truth. But when $\mathbb{P}_t$ gets  close to the boundaries, an observer will begin to observe more mean reverting values than she would consider to be statistically plausible. This anomalous behavior occurs when  $\mathbb{P}_0 < \Delta$ or $\mathbb{P}_0 > 1- \Delta$, where $\Delta$ is a thin sliver of width $\delta^2/2\lambda^3 = O(\lambda)$. Apart from these rare situations, an observer, using the R-estimator, would not be able to distinguish the small bias in her estimate from measurement noise.\footnote{In principle, this bias could be reduced by choosing a slightly larger value of $\lambda$, i.e. a slightly faster rule. But this increases the mean-square error of the estimator. The trade-off between the two would again lead to a negligible improvement of the bias of order $O(\lambda)$.}  

In conclusion, using a simple constant gain learning estimator leads to a negligible bias which is difficult or impossible to detect most of the time.  Of course, some smart agents could be aware that the death probability is non zero and account for it in their update rule. However, this would not necessarily make them more successful in the securities market that we will set up in the next section.\footnote{See the detailed discussion of this important point in section \ref{sec:Kelly} below.}

\section{Heterogeneous Beliefs in a Market Economy} \label{sec_hbelief}
We have built a model to describe the evolution of public opinion. But what  happens if  people  trade with other people with different beliefs? To answer that question we construct an endowment economy where each person is endowed with $\varepsilon$ units of a non-storable commodity in every period in which she is alive. We further assume that people  trade a complete set of Arrow securities, indexed to the {\em exogenous state}, which we represent by $\sigma$. We use the adjective {\em exogenous}, to distinguish the vector $\sigma$ from a vector of {\em endogenous states} that we introduce in Section \ref{later}.

We locate our agents in a market economy and we allow them to make trades on all publicly observable events. These events include, not only the binary signal that we refer to as public opinion, but also the realization of who lives and who dies in every period. This complication introduces $2^N$ new markets since every agent must, in a complete markets economy, trade life insurance contracts contingent on the survival of  everyone alive.

Much of Section \ref{sec_hbelief} involves the introduction of notation to deal with these additional life-insurance markets. Our main results refer to the large $N$ limit and, in this case, there is no aggregate risk from the mortality of individual agents. Although this means that the large $N$ results are much cleaner, we need the finite $N$ machinery to compute the difference between true probabilities and market-implied wealth weighted probabilities. It is this distinction, which does not disappear in the large-$N$ limit, that drives our main results.

\subsection{The Definition of the Exogenous State}
The  exogenous state has three elements.  The first element,  $s\in \mathbf{S} \equiv \{0,1\}$, is the realization of a public signal. The second element,  $\xx \in \mathbf{{X}} \equiv \{0,1\}^N$, is a vector that differentiates newborns from survivors and the third element, $\zz\in \mathbf{Z} \equiv [0,1]^N$, encodes the conditional probabilities of newborns.\footnote{We generate this vector for all $i$, including survivors from the previous period. Notice, however, that $z_i$ only enters the model when multiplied by $1-x_i$ which is zero for survivors.} Putting these pieces together we have that $\sigma \equiv \{s,\xx,\zz\} \in \mathbf{\Sigma} \equiv \mathbf{S} \times \mathbf{X} \times \mathbf{Z}$. We use a prime to denote the state in period $t+1$. 

At each date, people trade a complete set of Arrow securities which depend not just on $s'$, but also on the realizations of $\xx'$ which encodes who lives and who dies. There are $2^N$ possible realizations of $\xx'$ where the $i$'th element of $\xx'$ equals $\{1\}$ if person $i$ survives and $\{0\}$ if she dies. The $\sigma'=(s',\xx')$ security costs $Q(\sigma'|\sigma)$ commodities at date $t$  and pays $1$ commodity at date $t+1$ if and only if state $\sigma'$ occurs. We assume that everybody has different beliefs, represented by $\mathbb{P}_i(\sigma')$ that the state at period $t+1$ is $\sigma'=(s',\xx')$.  

This completes our definition of the exogenous state. In the subsequent subsection we define the objectives and constraints of individual agents and we derive a set of rules that represents their behavior in an exchange economy.

\subsection{A Model of Rational Choice}

We assume that  agents maximize the discounted expected utility of the logarithm of  consumption. This assumption implies that our agents choose to spend a fixed fraction of wealth in each period on the consumption good. The novel aspect of our approach is the decision rule we derive which shows how agents allocate their wealth to the two Arrow securities. This decision rule depends on their subjective beliefs, which evolve in the manner described in  Section \ref{Dis}.

First, we break wealth into two components; human wealth and financial wealth. The  {\em human wealth} of person $i$ is defined by the recursion,
\be \label{HWEQ}
H_i(\sigma) = \varepsilon +     \sum_{\sigma'} Q(\sigma'|\sigma) \, x_i' \,  H_i(\sigma'). 
\ee

Next, we define {\em financial wealth} of person $i$, $a_i(\sigma)$, to be the value of Arrow securities brought into period $t$. The   {\em total wealth} of person $i$ is the sum of human wealth and {\em financial wealth }
\be
W_i(\sigma) = H_i(\sigma) + a_i(\sigma).
\ee

Each period, the agent faces the following budget equation,
\be \label{E2}
\sum_{\sigma'} x_i'(\sigma') Q(\sigma'|\sigma) a_i'(\sigma') + c_i(\sigma) =  a_i(\sigma) + \varepsilon.
\ee
The right side of Eq.\ \eqref{E2} represents a person's available resources  at date $t$. The left side represents the ways those resource can be allocated;  to consumption or to the accumulation of  a bundle of Arrow securities that will be available for consumption or saving in the subsequent period.

We model the consumption and asset allocations of each person as the unique solution to the following maximization problem:
\begin{problem}\label{P1}
\be \label{E1}
V_i[W_i(\sigma)]=\max_{W_i'(\sigma')} \left[ \log c_i(\sigma) + \beta \sum_{\sigma'} \mathbb{P}_i(\sigma') x_i'(\sigma') V_i'[W_i'(\sigma')] \right]
\ee
such that
\begin{equation}\label{iteration3a} 
\mathbb{P}_i(\sigma') = x'_i[ (1 - \lambda) \mathbb{P}_i(\sigma) + \lambda s] + (1-x'_i) z'_i, 
\end{equation}
and
\be \label{E3}
\sum_{\sigma'} x_i(\sigma')Q(\sigma'|\sigma) W_i(\sigma') + c_i(\sigma) \leq  W_i(\sigma) .
\ee

\end{problem}
In Section \ref{Dis} we derived an expression for the evolution of person $i$'s beliefs. Eq.\ \eqref{iteration3a} reproduces that equation using the definition of $\sigma$ and replacing time subscripts with prime notation. 

Eq.\ \eqref{E3} is derived by combining equations \eqref{HWEQ} and \eqref{E2} with the assumption that agents must remain solvent.  $V_i[W_i(\sigma)]$ is the maximum attainable utility given wealth $W_i(\sigma)$, $c_i(\sigma)$ is date $t$ consumption and $\beta$ is the common discount rate. Following common usage we refer to the consumption decision that solves Problem \ref{P1} as the {\em policy function} and to the maximum attainable utility as a function of wealth as the {\em value function}.

\begin{proposition} \label{Prop1}
The policy function and the value function for  Problem \ref{P1} are given by  Equations \eqref{E4} and \eqref{E5},
\begin{align} \label{E4}
c_i(\sigma) & = [1-\beta(1-\delta)] W_i(\sigma), \\ \nonumber \\
V_i[W_i(\sigma)] & = \frac{1}{1 - \beta(1-\delta)} \log[W_i(\sigma)] + B, \label{E5}
\end{align}
where $B$ is a constant that can be computed but its value is irrelevant for our purpose.

The wealth of the person with label $i$ evolves according to  Eq.\ \eqref{E6}
\be \label{E6}
W_i(\sigma')  = x'_i  \left[\frac{\beta \mathbb{P}_i(\sigma')}{Q(\sigma'|\sigma)} W_i(\sigma) \right] + (1-x'_i) H_i(\sigma'),
\ee
where $H_i(\sigma)$ is defined by the recursion Eq.\ \eqref{HWEQ}.
\end{proposition}

The first term on the right side of Eq.\ \eqref{E6} is the wealth evolution equation for person $i$ if she survives into period $t+1$. The second term on the right side of the equation resets person $i$'s wealth  to $H_i(\sigma')$ if she dies and is replaced by a newborn. For a proof of Proposition \ref{Prop1}, see Appendix \ref{appsolve}. 

\subsection{Definition of Equilibrium} \label{later}

We have constructed a theory of individual choice. According to this theory, peoples' decisions are a function of the exogenous state and of the stochastic process for prices. In this section we construct an equilibrium  theory where  prices are determined  by setting the excess demands for goods and the excess demands for  Arrow securities, in every period, to zero. First, we define a new object;  the {\em endogenous state}.

The endogenous state  has two elements. The first element,  $P \in \mathbf{P} \equiv [0,1]^N  $ is a vector of subjective conditional probabilities with generic element $\mathbb{P}_i$.  The second element, $W \in \mathbf{W} \equiv \mathbb{R}_+^N$ is a vector of wealth positions with generic element $W_i$. Putting these pieces together, the endogenous state is represented by $y\equiv \{P,W\} \in \mathbf{Y} \equiv \mathbf{P} \times \mathbf{W}.$

Next, we derive a function $\mathcal{G}(\cdot)$ to explain how the endogenous state evolves through time. Our approach is a relatively standard application of recursive equilibrium theory \citep{StokeyLucasPrescottBook}. Our innovation, over conventional dynamic stochastic general equilibrium models,   is to provide a  self-referential theory of learning in which the economy does not converge to a rational expectations equilibrium. 

We begin with a definition of recursive equilibrium:

\begin{definition}[Recursive Equilibrium]
A  recursive equilibrium is a price function $Q: \mathbf{\Sigma}^2 \rightarrow \mathbf{Q} \equiv [0,1]^{2N}$ and a state evolution function $\mathcal{G}: \mathbf{Y} \times \mathbf{\Sigma} \times \mathbf{Q} \rightarrow \mathbf{Y} $ with the following properties:
\begin{enumerate}
    \item The state evolution function, $\mathcal{G}$, is given by equations \eqref{iteration3a} and \eqref{E6}. This function determines the evolution of the vector of beliefs, $P$, and the vector of wealth positions, $W$.
    \item When the Arrow security prices are given by $Q(\sigma'|\sigma) $ and when $y'=\mathcal{G}(y;\cdot)$ the implied consumption plan solves Problem \ref{P1}.
    \item The goods market clears for all $\sigma'$ where $c_i(\sigma')$ solves Problem \ref{P1}: 
    \be 
    \sum_{i=1}^N c_i(\sigma') = N \varepsilon.
    \ee 
    \item The Arrow securities markets clear for all $\sigma'$ where $a_i(\sigma') = W_i(\sigma') - H_i(\sigma')$: 
    \be 
    \sum_{i=1}^N a_i(\sigma') = 0.
    \ee
\end{enumerate}
\end{definition}

In Proposition \ref{Prop2}, we show that, in equilibrium, human wealth is a number that does not depend on the state and we derive an expression for the equilibrium price function $Q(\sigma'|\sigma)$.

\begin{proposition} \label{Prop2}
In a recursive equilibrium:
\begin{enumerate}
    \item Individual human wealth $H_i$ is independent of $\sigma$ and is the same for all agents. It is given by the expression,
    \be \label{HW}
    H = \frac{\varepsilon}{1-\beta(1-\delta)}.
    \ee
    \item The price of an Arrow security is given by Eq.\ \eqref{Qeqb},
    \be \label{Qeqb}
    Q(\sigma'|\sigma) = \beta  \frac{\sum_{i=1}^N  \mathbb{P}_i(\sigma') x_i' W_i(\sigma)}{N(\sigma') H},
    \ee 
    where $N(\sigma')=\sum_i x'_i$ is the number of surviving agents at time $t+1$ and  $N(\sigma') H $ is aggregate human wealth.
\end{enumerate}
\end{proposition}
For a proof of Proposition \ref{Prop2} see Appendix \ref{appsolve1}. In the next section, we will show how the pricing function, $Q(\sigma'|\sigma)$, depends on the assumptions about the information structure and the number of agents. 

\section{Equilibrium Behavior Under Two Different Assumptions}
\label{EBU}
Next, we  study the evolution of asset prices and  the wealth distribution under two different assumptions.  First, in Section \ref{CK}, we assume that $\mathbb{P}_i(\sigma')=\mathbb{P}(\sigma')$ for all $i$ We call this the {\em common knowledge economy} and  we refer to the  outcome of this version of our model as a rational expectations equilibrium.  In Section \ref{HB} we allow beliefs to differ and we ask and answer the question: Do markets reveal enough information for the economy to converge to a rational expectations equilibrium? We call this the {\em heterogeneous beliefs economy}.

\subsection{The Common Knowledge Economy} \label{CK}
When beliefs about the probability of $s'$ are common, and $s'$ and $\xx'$ are independent,  we can write Eq.\ \eqref{Qeqb} for $Q(\sigma'|\sigma)$  as follows,
\begin{align} \label{Qit}
Q(\sigma'|\sigma)  & = \beta \mathbb{P}(s')p(\xx')\theta(\xx'),
\end{align}
where, $p(\xx')$ is the commonly held probability for the vector of survival outcomes $\xx'$ and $\mathbb{P}(s')$ is the commonly held probability for the next state being $s'=1$ or $s'=0$. Furthermore we have used  the equality $\sum_{i=1}^N x'_i=N(\sigma')$ to define the variable $\theta(\xx')$ as follows,
\[
 \theta(\xx') = 1 + \frac{\sum_{i=1}^N a_i(\sigma) x'_i}{N(\sigma') H}.
\]
The term $\theta(\xx')$ corrects Arrow security prices for mortality risk and we need to keep track of this term in our simulations to ensure that asset markets clear. This term disappears in the large $N$ limit because each cohort is perfectly insured. As $N\rightarrow \infty$,  $\theta(\xx') \rightarrow 1$ and we obtain the limiting expression\footnote{Notice that $   \plim_{N \rightarrow \infty} N^{-1} \sum_i a_i(\sigma) x_i'=0,$ using market clearing and assuming that  $\plim_{N \rightarrow \infty} N^{-2} \sum_i a_i^2 =0$, which turns out to be true provided $\delta$ remains fixed as $N \to \infty$. Hence $\plim_{N \rightarrow \infty} \theta(\xx') = 1.$}
\be \label{E21}
Q(\sigma'|\sigma) = \beta \mathbb{P}(s')p(\xx').
\ee

Consider next the following expression for the evolution of wealth over time which follows from from Eq.\ \eqref{E6}, Eq.\ \eqref{Qit}  and the fact that $H$ is state independent and common to all individuals,
\be 
W_i(\sigma')  = x'_i  \left[ \frac{W_i(\sigma)}{\theta(\xx')}\right]  + (1-x'_i) H.
\ee 
In the large $N$ limit, there is no aggregate mortality risk and, in this case, we obtain the following  expression for $W_i(s)$
\be \label{Ws}
W_i(s')  = x'_i  W_i(s)  + (1-x'_i) H.
\ee 
Eq.\ \eqref{Ws}  implies that in the large $N$ economy, the wealth of the person with index $i$, contingent on her survival, is time invariant.

In a finite population, the variable $\theta(\xx')$ plays a non-trivial role. Suppose, for example, that in period $1$ there are two people. One person  has positive financial assets equal to $a$ and the other has negative financial assets equal to $-a$. In that economy, the rich person consumes more than the poor person for as long as they are both alive. But if one person dies and is replaced by a new person with wealth $H$,  all debts are canceled and the economy enters an absorbing state  with an egalitarian wealth distribution. The wealth reallocation that occurs as a consequence of mortality risk is encoded into the random variable $\theta(\xx')$.  

\subsection{The Heterogeneous Belief Economy}\label{HB}
Next, we turn to the case where people have different beliefs. In this case,  $\mathbb{P}_i(\sigma')$ can no longer be factored out of the summation in Eq.\ \eqref{Qeqb} and instead of Eq.\ \eqref{Qit} we obtain the following expression for the  price of an Arrow security,
\begin{equation}
\label{eq_full_implied}
    Q(\sigma'|\sigma) = \beta p(\xx')  \left(  \frac{ \sum_{i=1}^N\mathbb{P}_i(s') W_i(\sigma) x'_i} {N(\sigma') H } \right)\equiv \beta  \mathbb{P}_{\text{imp}}(\sigma') p(\xx'),
\end{equation}
where $\mathbb{P}_{\text{imp}}(\sigma')$ is defined as the probability of state $\sigma'$ that would be inferred from market prices if market participants believed that they were living in a common knowledge economy. We henceforth refer to $\mathbb{P}_{\text{imp}}(\sigma')$ as the {\it implied probability}.

Since who dies and who survives is independent from both wealth and beliefs one has, in the large $N$ limit,\footnote{We use here the fact that if $\eta_i$ and $\xi_i$ are independent random variables, then
\[
\plim_{N \to \infty} N^{-1} \sum_{i=1}^N \eta_i \xi_i = \plim_{N \to \infty} \left(N^{-1} \sum_{i=1}^N \eta_i \right)
\left(N^{-1} \sum_{i=1}^N \xi_i \right),
\]
and choose $\eta_i=\mathbb{P}_i W_i$ and $\xi_i=x_i'$.}
\be \label{QP2sig}
\plim_{N \to \infty} \left( \frac{\sum_{i=1}^N\mathbb{P}_i(s') W_i(\sigma) x_i'}{N(\sigma')H} \right)  = \plim_{N \to \infty} \left( \frac{\sum_{i=1}^N\mathbb{P}_i(s') W_i(s)}{NH}\right),
\ee 
where we distinguish $N$, which refers to the number of people in state $\sigma$ at date $t$, from $N(\sigma')$, which is the number of survivors in state $\sigma'$ at date $t+1$. 

In the large $N$ limit, $\mathbb{P}_{\text{imp}}(\sigma')$ depends on the future realisation of $s$ but not on the mortality state. It  is given by the expression,
\begin{equation}\label{implied_P}
    \mathbb{P}_{\text{imp}}(s') \equiv    \frac{ \sum_{i=1}^N\mathbb{P}_i(s') W_i(s)} { N H }.
\end{equation}
$\mathbb{P}_{\text{imp}}(s')$ is the wealth weighted average probability and it differs from the true probability, $\mathbb{P}(s')$, which is the unweighted average of individual subjective probabilities.

Using the definition of $\mathbb{P}_{\text{imp}}(s')$, the analogue of Eq.\ \eqref{Ws} for the heterogeneous belief case  is given by Eq.\ \eqref{eq_full_implied_2},
\be \label{eq_full_implied_2}
W_i'(s')=  x_i' \frac{\mathbb{P}_i(s')}{\mathbb{P}_{\text{imp}}(s')} W_i(s)  + (1-x_i') H.
\ee
$\mathbb{P}_i$ and $W_i$ are {\em strongly coupled} by the dynamics of individual wealth accumulation, Eq.\ \eqref{eq_full_implied_2},  and because of this strong coupling  we cannot split $\mathbb{P}_\text{imp}(s')$ into the product of $\mathbb{P}(s')$ and $\plim_{N \rightarrow \infty}{\sum_{i}(W_i(\sigma)}/N)$, even asymptotically,  as we did in the common knowledge economy.
This failure of independence generates fat-tails in the wealth distribution and it implies that the implied probability,  $\mathbb{P}_{\text{imp}}(s')$, and the true probability, $\mathbb{P}(s')$,  can  differ {\em even in the large $N$ limit}.\footnote{We explore the implications for the wealth distribution in Section \ref{wealth}.}

\subsection{Debt and Equity in the Heterogeneous Belief Economy} \label{DEq}
 We have  derived explicit  trading rules for agents who buy and sell Arrow securities.  But there is no reason to restrict ourselves to securities of this kind and the same equilibrium we  described above can be supported by any set of  securities with independent payoffs that span the space of possible outcomes.  In this subsection we show that, in the large $N$ limit,  an equilibrium can be supported by a security that pays one commodity in both states; we call this security {\em debt}, and a security that pays $d$ units  if $s=\{1\}$ and zero otherwise; We call this security {\em equity}.
 
The assumption that $N$ is large allows us to ignore fluctuations in the annuities markets and to concentrate on trades contingent on disagreement over the realization of $s'$. This signal could be any mechanism for the revelation of public opinion. What is important for our interpretation is that firms choose to pay dividends only if $s'=1$. 

\begin{proposition}\label{prop3}
For the large $N$ economy, equilibrium can be supported by trades in debt and equity. Debt is a security that costs $Q_t$ units of commodities at date $t$ and pays $1$ commodity at date  $t+1$ in both states. Equity is a security that costs $p_{\text{E},t}$ units of commodities at date $t$ and pays $d$ in state $s'_{t+1}=\{1\}$ and $0$ in state $s'_{t+1}=\{0\}$, where

\be \label{eq_equity}
p_{\text{E},t} = \frac{d \beta}{2} \left[\frac{2\mathbb{P}_{{\text{imp},t}}-1}{1-\beta(1-\delta)} + \frac{1}{1-\beta} \right], 
\ee 
\be
Q_t = \beta.
\ee
\end{proposition}
For a proof of Proposition \ref{prop3} see Appendix \ref{proof3}.


\subsection{What Drives Our Results?}
In the work of \citet{bek-esp-1},  agents all eventually agree. We make two deviations from the Beker-Espino environment. First, we assume that agents die and are replaced by new agents and second, we assume that the process that agents learn about is both self-referential and non-fundamental. 

The assumption that people die 
is essential to our result that the economy fails to converge to a rational expectations equilibrium. The assumption that the stochastic process for $s_t$ is self-referential is secondary but important. The defining property of this process is that of quasi-non-ergodicity.  

If agents never die, there is no advantage to the use of  constant-gain learning. If, as would be optimal in the infinite-lives environment, everyone were to use least-squares learning,  agents would converge asymptotically to the truth and they would all agree, as in Beker and Espino, although convergence would be slow, and the final belief would be history dependent. 

One {\em could} assume that the process for $s_t$ is exogenous but quasi-non-ergodic. It might, for example, be generated by the equation,
\begin{equation}\label{diffproc}
\mathbb{P}_{t+1} = (1-\delta_1)\left[(1 - \lambda_1) \mathbb{P}_{t} + \lambda_1 s_{t}\right] + \frac{\delta_1}{2}, 
\end{equation}
where $\lambda_1$ and $\delta_1$ are not necessarily equal to $\lambda$ and $\delta$. If agents were to learn about this process using constant gain learning with gain parameter $\lambda=\lambda_1$, and if the parameter $\delta_1$ were by chance, equal to the death probability $\delta$,  the model in which the stochastic process for $\mathbb{P}$ is exogenous would be indistinguishable from the model we have presented here. 

It is plausible that a three-parameter model in which $\delta$ and $\delta_1$ were different but where everybody had correctly learned the parameter $\lambda_1$, would display similar behavior to the model we have described here. We have not explored that variant of our main theme. The importance of our interpretation  of Eq. \eqref{diffproc}, as a self-referential process driven by social interactions, is that it provides a micro-founded theory for the assumption that agents must learn about a quasi-non-ergodic process. In the absence of our interpretation of this equation one would need to find some  alternative economic explanation for what we think is an attractive feature of our work: the idea that the value of stocks today  depend on guessing what others think that stocks will be worth in the future. In our model, the interaction of market and non-market forces generates a micro-founded model of why people disagree that has implications for both the wealth distribution and excess volatility in the asset markets.  Both of those implications are pursued further below.

\section{Results from Simulated Data}\label{Sim}
In this Section  we illustrate the implications of our results by reporting some statistics for simulated data in a calibrated version of our model. 

\subsection{A numerical simulation}
\label{sec_simulation}

We simulated an economy with one million agents for 300 years and we chose the period length to be one week. We normalized the weekly endowment to 1 and we chose the annual discount rate to be $0.97$ which corresponds to an equilibrium annual real interest rate, in an endowment economy, of 3\%. These are relatively uncontroversial choices. 

\begin{figure}[!thb] 
\begin{center}
\includegraphics[width = \textwidth]{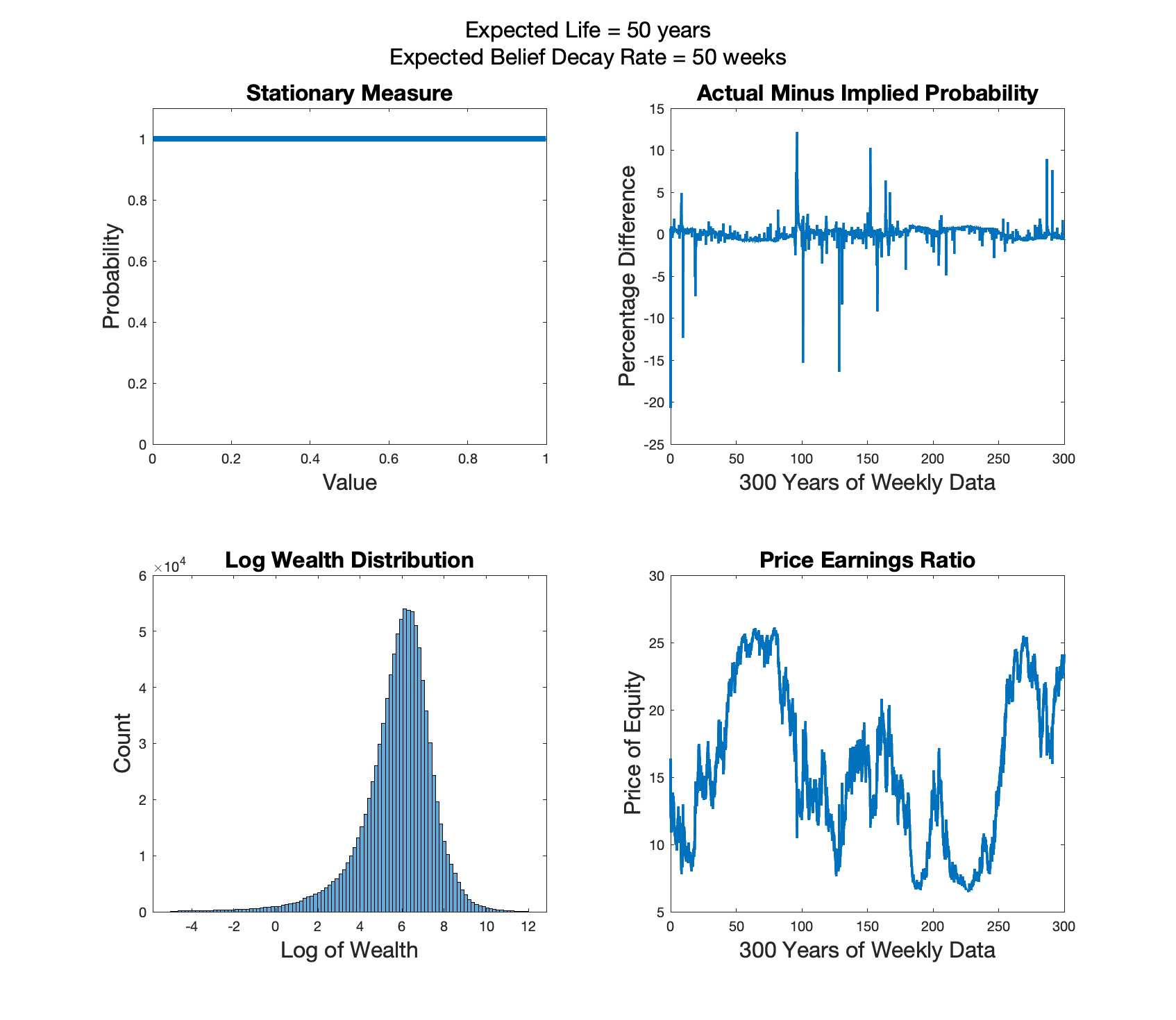}
\caption{300 Years of Simulated Weekly Data in an Economy with One Million People}\label{fig1}
\end{center}
\end{figure}

In Figure \ref{fig1} we graph some  data from a single simulation of this calibrated version of our model when agents have a life-expectancy of $50$ years and for a memory time of one year. Importantly, for our calibration, $\alpha=1$ and the invariant measure is uniform. We checked that our reported results, especially those concerning the wealth distribution, are robust to values of $\alpha \in [0.5,2]$, see e.g. Fig. \ref{Fig3} for the level of disagreement generated by the model. We suspect that our main results are relatively insensitive to the choice of $\alpha$ even outside of this range.\footnote{Only by taking  $\alpha$ to $10$, were we able to break one of our key results; that the wealth distribution has fat tails.}  


The top left panel of Figure \ref{fig1} graphs the invariant measure $\mathcal{P}_\infty(\mathbb{P})$. The other three panels present some key data for a single simulation of 300 years of weekly data. The top right panel is the percentage difference between $\mathbb{P}(s')$ and $\mathbb{P}_{\text{imp}}(s')$. This difference is a measure of how wrong the market can be as a measure of the true probability. For much of the sample this difference is less that $1\%$ but there are times when this deviation exceeds ${\pm}15\%$. Such large discrepancies are quite remarkable in view of the size of the market (one million participants) and are the consequence of the emergent wealth inequalities in our model.  

The bottom right panel shows the time series behavior of the price-earnings ratio using the formula derived in Section \ref{DEq}. To compute this series we normalized the dividend payment to 1/52 to make the units comparable to an expected weekly dividend payment. This series has many characteristics in common with the price dividend ratio in US data for realized values of the S\&P. It wanders randomly over a bounded interval and sometimes it moves substantially in a short period of time. The bottom left panel is the distribution of the log of wealth. In Section \ref{wealth}, we explore the properties of this distribution further and we show that it shares many characteristics in common with empirical wealth distributions in Western economies.

\section{Exploring the Empirical Wealth Distribution}\label{wealth}

While our model is constructed in such a way that no agent is better informed than any other, some agents are temporarily, purely by chance, much more successful than others. This allows these agents to accumulate wealth through the multiplicative process described in Eq.\ \eqref{eq_full_implied_2}, reproduced below
\be \tag{\ref{eq_full_implied_2}} \label{eq_full_implied_21}
W_i'(s')=  x_i' \frac{\mathbb{P}_i(s')}{\mathbb{P}_{\text{imp}}(s')} W_i(s)  + (1-x_i') H.
\ee
Multiplicative wealth processes of this form  are well-known to generate important wealth inequalities.\footnote{\label{f1} Examples of papers in the literature that study multiplicative wealth dynamics include  \citet{kesten, bouchaud_mezard,ben-bis-1,gabaix_tails,ben-bis-zhu-1} and \citet{gabaix}.}
\begin{figure} 
\centering
\includegraphics[width = \textwidth]{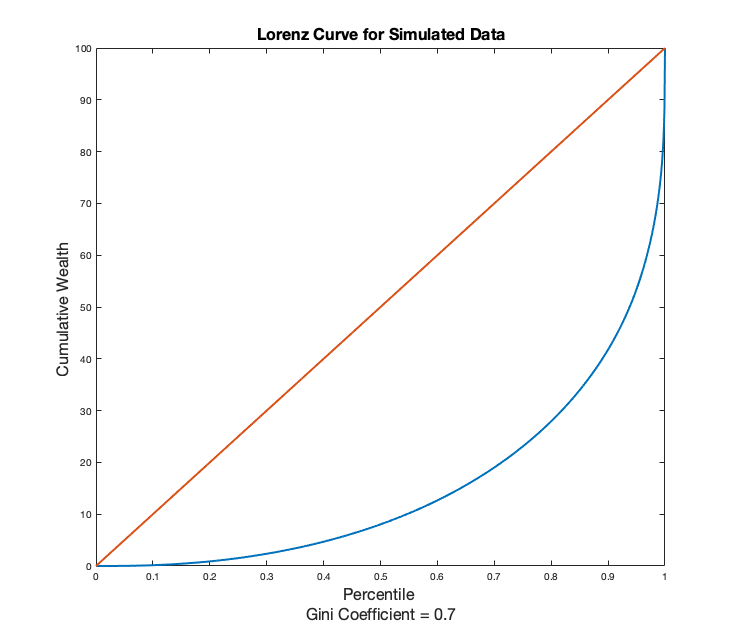}
\caption{The Lorenz Curve for a Single Simulation}
\label{fig3}
\end{figure}

In Figure \ref{fig3} we graph the Lorenz curve for the time average of 250 equally spaced samples of the wealth distribution in our simulated data.\footnote{For large $T$, our sample histogram will converge to the ergodic wealth distribution. There will still be some  variability in a sample of 300 years but our experiments with different random draws suggest that this variability not too large.}   The Lorenz curve is a  graphical representation of  inequality which plots the cumulative percentage of wealth on the y-axis against the percentile of the population on the x-axis. One popular index of inequality is the Gini coefficient which is equal to  twice the area between the 45 degree line and the Lorenz curve.

For our numerical data, the Gini coefficient is equal to  $0.7$. A value of 0 would represent a completely equal distribution and a value of 1 would represent a distribution where one person owns everything. 
Table \ref{table1} reports data from a selection of countries. This table shows that a Gini coefficient of $0.7$ is well within the bounds of empirical data which varies between a low of $0.55$ for China in 2008 and a high of $0.85$ for the United States in 2019.
\begin{table}[tbh!]
\begin{center}
\begin{tabular}{ l | l | l }
   Country   &  2008 & 2019  \\
   \hline \hline
   China & $0.55$ & $0.7$ \\
   United Kingdom   & $0.7$ & $0.75$ \\
   Italy & $0.7$ & $0.77$ \\
   France & $0.73$ & $0.7$ \\
   Switzerland & $0.74$ & $0.87$ \\
   United States & $0.8$ & $ 0.85$ \\
   \hline
\end{tabular}
\caption{Wealth Ginis' For a Selection of Countries in 2008 and 2019\protect\footnotemark \label{table1}}
\end{center}
\end{table}
 \footnotetext{Wikipedia https://en.wikipedia.org/wiki/List\_of\_countries\_by\_wealth\_equality Retrieved December 6'th 2020.}
 
\begin{figure}
\centering
\includegraphics[width = \textwidth]{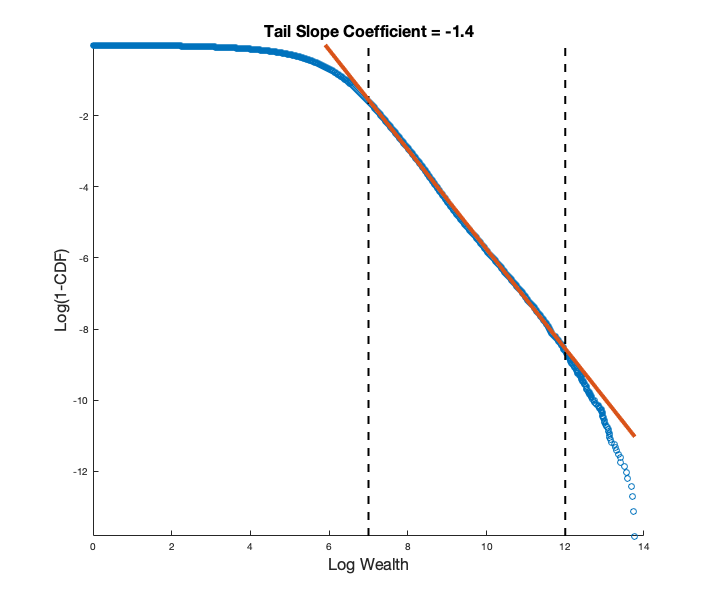}
\caption{Estimate of the Tail Parameter in 1,000 Years of Simulated Monthly Data}
\label{fig2}
\end{figure}

 To explore the nature of the wealth distribution further we define $F(W)$ to be the cumulative distribution function (cdf) of wealth and define $G(W) \equiv 1-F(W)$ to be the complementary cdf. In Figure \ref{fig2}, we plot $\log G(W)$ against  $\log(W)$ for values of $\log(W)$ greater than zero. This figure reveals  a  power-law tail of the form $G(W) \sim W^{-\mu}$, and a regression of  $\log(G(W))$ on $\log(W)$ for the linear portion of the plot  provides an estimate of the tail index of  $\mu=1.4$. Note that $G(W) \sim W^{-\mu}$ corresponds to a probability distribution function (pdf) $\varrho(W) \sim W^{-1-\mu}$.
 A person who is neither a borrower nor a lender has zero financial assets and her net worth would be equal to the discounted present value of her labour income. For our calibration, this number, which we refer to as human wealth, is equal to  1,032 weeks of income.\footnote{ Human wealth is defined by the expression $H = 1/(1-\beta(1-\delta))$. For our calibration the weekly discount rate is $0.97^{1/52}$ and the survival probability, $(1-\delta)$, is equal to $3.9\times 10^{-4}.$ This leads to a value of $H=1,031$ measured in weeks of income.} In the common knowledge economy, the wealth distribution would be egalitarian, the Gini coefficient would be $0$ and everyone would have wealth equal to $H$. Instead, in our economy, there is considerable inequality.
 
 A person at the $50$'th percentile of the wealth distribution is a net borrower who has  total wealth equal to $39\%$ of human wealth.  In contrast, a person at the $99'th$ percentile in the wealth distribution has total wealth equal to $892\%$ of human wealth and the person at $99.9'th$ percentile has total wealth of $4,999\%$. Wealth becomes highly concentrated because  market prices do not reflect average beliefs. Instead they reflect wealth weighted beliefs.  In equilibrium, wealth and market prices are correlated  in a way that leads to a self-reinforcing mechanism whereby a few people, by chance, get lucky and become very rich. 
 
 To explore the dependence of our results on $\alpha$ we recomputed the data reported in Figure \ref{fig1} for two alternative values of $\alpha$. For both simulations we held $\delta$ constant and selected values of  $\lambda$ that set $\alpha$ to $0.5$ and $2$.  For $\alpha=2$, we find a Gini coefficient of $0.69$ and  a tail slope coefficient of $1.4$. This calibration has a hump-shaped invariant measure and a memory time of $72$ weeks. For $\alpha=0.5$ we find a Gini coefficient of  $0.72$ and once again, a tail slope coefficient, to one significant digit, of  $\mu=1.4$. This calibration has a U-shaped invariant measure and a memory time of $36$ weeks. We infer from these robustness checks that  our results are insensitive to variations in $\alpha$ for a substantial range of plausible parameters. In fact, as we explain below,  $\mu$ converges to a non-degenerate value $ > 1$ when $\delta \to 0$.
 
\subsection{The Behavior of Wealth in the Large $N$ Limit} \label{sec_limit}

We can learn quite a bit about the dynamics of wealth by analyzing the properties  of Eq.\ \eqref{eq_full_implied_2}. Using  this equation, one may derive the following   expression for the  average return for agent $i$ between dates  $t$ and $t+1$, conditional on surviving:\footnote{Eq.\ \eqref{eq_return} follows since \begin{align*} 
R_i\equiv \mathbb{E}\left[\frac{W'_i}{W_i}-1\right] &= \left[\mathbb{P} \times \frac{\mathbb{P}_i}{\mathbb{P}_{\text{imp}}} + (1-\mathbb{P}) \times \frac{1-\mathbb{P}_i}{1-\mathbb{P}_{\text{imp}}} - 1 \right]\\ 
&= \frac{(\mathbb{P}-\mathbb{P}_{\text{imp}})(\mathbb{P}_i-\mathbb{P}_{\text{imp}})}{\mathbb{P}_{\text{imp}}(1-\mathbb{P}_{\text{imp}})}.
\end{align*}}

\begin{align} \label{eq_return}
R_{i}\equiv \mathbb{E}\left[\frac{W'_{i}}{W_{i}}-1\right] = \frac{(\mathbb{P}-\mathbb{P}_{\text{imp}})(\mathbb{P}_{i}-\mathbb{P}_{\text{imp}})}{\mathbb{P}_{\text{imp}}(1-\mathbb{P}_{\text{imp}})}.
\end{align}
Several interesting conclusions can be drawn from Eq.\ \eqref{eq_return}. First, in the common knowledge economy where $\mathbb{P} \equiv \mathbb{P}_{\text{imp}}$, agents cannot expect to make money on average, even temporarily. 

Second, when an agent's belief $\mathbb{P}_i$ is larger than the market probability $\mathbb{P}_{\text{imp}}$, her expected gain is positive if the actual probability  $\mathbb{P}$ is also greater than $\mathbb{P}_{\text{imp}}$, and negative otherwise. In fact, provided the sign of $\mathbb{P}_i-\mathbb{P}_{\text{imp}}$ is the same as that of $\mathbb{P}-\mathbb{P}_{\text{imp}}$, the instantaneous expected gain is larger when the bet is bolder, albeit with a larger variance (see Eq. \eqref{eq_variance} below). 

Finally,  since agents are assumed to act on the assumption that their estimate of the probability is an unbiased estimate of the true probability, they also believe that their trades will be profitable on average and proportional to $(\mathbb{P}_i-\mathbb{P}_{\text{imp}})^2$. In other words, they expect to make a larger profit, the further is their belief from   the probability implied by the market price. This implies that there is no incentive for agents to align their beliefs with  the observable implied probability, since this would reduce their subjective expected profit. Everybody in this economy, believes that they know more than the market -- indeed, a most common feature of the real world! 

In Eq.\ \eqref{eq_variance} we derive an expression for  the average of the square of the relative change of wealth for surviving agents:\footnote{Eq.\ \eqref{eq_variance} follows from 
\begin{align*}
 \mathbb{E}\left[\left(\frac{W'_{i}}{W_{i}}-1\right)^2\right] &= \left[\mathbb{P} \left(\frac{\mathbb{P}_{i}}{\mathbb{P}_{\text{imp}}}-1\right)^2 + (1-\mathbb{P}) \left(\frac{1-\mathbb{P}_{i}}{1-\mathbb{P}_{\text{imp}}} - 1 \right)^2 \right].
\end{align*}.}
\begin{equation}
 \mathbb{E}\left[\left(\frac{W'_{i}}{W_{i}}-1\right)^2\right] = \frac{\left(\mathbb{P}(1- \mathbb{P}_{\text{imp}})^2+(1-\mathbb{P})\mathbb{P}_{\text{imp}}^2\right)(\mathbb{P}_{i}-\mathbb{P}_{\text{imp}})^2}{\mathbb{P}_{\text{imp}}^2(1-\mathbb{P}_{\text{imp}})^2}.\label{eq_variance}
\end{equation}

One sees from this equation that ``bold beliefs'', corresponding to a large difference between $\mathbb{P}_i$ and the market probability $\mathbb{P}_{\text{imp}}$, leads to a larger variance of gains. Eq.\ \eqref{eq_variance}  explains why our model generates  large wealth inequalities. For surviving agents, the wealth dynamic is a multiplicative random process with a time dependent and agent dependent variance. This multiplicative process is  reset to 1 at a Poisson rate $\delta$, i.e. when an agent dies. 

Multiplicative random process with reset have been widely studied in the literature\footnote{See the citations in footnote \ref{f1}.} and it is known that such processes lead to a stationary distribution with a power-law tail with a pdf $\varrho(W)$ and  a complementary cdf $G(W)$ of the form,
\begin{eqnarray}\label{eq_power_law}
\varrho(W) \sim_{W \to \infty} W^{-1-\mu}, \qquad \qquad 
G(W) \sim_{W \to \infty} W^{-\mu}, 
\end{eqnarray}
where the exponent $\mu$ depends on the parameters of the problem.\footnote{Random variables with a Pareto tail can be sorted into three classes depending on the value of  the tail parameter $\mu$. A Pareto-tailed distribution is well defined for all positive $\mu$ but when $0<\mu\leq1$, the mean and all higher moments do not exist. When $1<\mu \leq 2$, the mean exists but the variance and higher moments do not exist and for $\mu>2$, the distribution  has a finite mean and a finite variance. In our example, as in the data, we find a value of $\mu$ between 1 and 2 which implies that the wealth distribution has a finite first moment but all higher order moments are not well defined.}  We discuss in Appendix \ref{multiplicative} how $\mu$ can be approximately computed and we find, in particular, that $\mu > 1$ whenever $\delta > 0$.\footnote{The limit $\delta \to 0$ is interesting since  the wealth distribution has a Pareto Tail  even as $\delta \to 0$. Using Eq. \eqref{eq_power_law2} in Appendix F, and taking $\delta$ to $0$ for fixed $\lambda$, the Pareto exponent $\mu$ converges to $\frac{1}{2}\left(1 + \sqrt{1+8\lambda}\right)$ which equals $1.23$ when $\lambda = 0.14$. In the limit, two effects cancel each other out as the limit of $\frac{\delta}{\sigma^2}$ converges to $1$. Disagreements tend to disappear (see Eq. \eqref{eq_dispersion}) and thus mispricings vanish (i.e. $(\mathbb{P}_{\text{imp}} - \mathbb{P})^2 \sim \delta$), but at the same time lucky agents can benefit from these mispricings for a longer time and $T \sim \delta^{-1}$.}

In conclusion, wealth inequalities in our model arise from the multiplicative nature of wealth dynamics which makes successful bold bets highly profitable. Unsuccessful bold bets, however,  are ruinous and lead the person who makes such bets into poverty. People who agree with the market belief have a low expected subjective  gain from trading.  People who disagree may either become spectacularly rich, or spectacularly poor. 

\subsection{The Kelly Criterion}\label{sec:Kelly}

In Section \ref{review}, we discussed the market selection hypothesis which is the claim that the agents who survive will be those who hold beliefs that are closest to the  truth. It is equivalent to the assertion that those agents who dominate the asset markets will be those who maximize the growth of their wealth and the \cite{blu-eas-1} formulation of this hypothesis implies that all surviving agents will hold common beliefs that converge asymptotically to the rational expectation.

The investment strategy that maximizes the growth rate of wealth was studied by \citet{joh-kel-1} and it is widely referred to as the Kelly criterion.   In the context of our model, the Kelly criterion amounts to maximizing the quantity $\mathbb{E}[\log W_i'/W_i]$. We seek an approximation to this quantity that is valid when the degree of disagreement, $\upsilon \equiv \mathbb{V}[\mathbb{D}_{i,t}]$, is small. This term is defined by Eq.\ \eqref{eq_dispersion},
\be\tag{\ref{eq_dispersion}}
\upsilon \equiv \mathbb{V}[\mathbb{D}_i] = \left[\frac{\lambda}{2 + ( \alpha-1)\lambda}\right] \, \frac{\alpha(\alpha +2)}{6(2\alpha +1)} \, + O(\lambda^3), 
\ee 
and for $\upsilon$ much smaller than unity, we have the following expansion for  the log of the growth rate of agent $i$'s wealth,
\begin{equation}\label{e40}
    \log \left(\frac{W_i'}{W_i} \right) \approx \left(\frac{W_i'}{W_i} - 1\right) - \frac12 
    \left(\frac{W_i'}{W_i} - 1\right)^2 + \ldots.
\end{equation}
Taking expectations of Eq.\ \eqref{e40}, using equations \eqref{eq_return}, and \eqref{eq_variance}, one finds the following approximation to the first order in $\upsilon$,
\begin{equation} \label{eq:Kelly}
    \mathbb{E}\left[\log \frac{W_i'}{W_i}\right] \approx  \frac{\left[(\mathbb{P}-\mathbb{P}_{\text{imp}})(\mathbb{P}_i-\mathbb{P}_{\text{imp}})- \frac12 
    (\mathbb{P}_i-\mathbb{P}_{\text{imp}})^2 \right]}{\mathbb{P}_{\text{imp}}(1-\mathbb{P}_{\text{imp}})}  + o(\upsilon).
\end{equation}
Eq.\ \eqref{e40} implies that when the market implied probability 
$\mathbb{P}_{\text{imp}}$ is equal to the true probability $\mathbb{P}$, any belief $\mathbb{P}_i \neq \mathbb{P}$ leads to a negative growth rate for $W_i$. Any  agent $i$  who continues to hold an inaccurate belief of this kind will be  wiped out in the long run. This is the content of the market selection hypothesis. 

In our model the entry of new agents causes the market selection hypothesis to fail.  It is the wealthy agents who determine the market price and, even though these agents would be wiped out if everyone lived forever, in the finite lived environment  the smart agents die before they have had time  to benefit from their more accurate subjective beliefs. The richest agents in any given period are not the smartest ones but rather those who have made bold  successful bets.  The market price is determined by rich people who were right in the past whereas those people with beliefs that are closer to the current truth will only become rich in the future.


\section{Conclusion} 

We have a constructed a  theory of beliefs in which people exchange information through both market and non-market interactions. Non-market interaction generates  an aggregate signal which reflects average public opinion. Market exchange through the purchase and sale of financial assets allows people to bet on their beliefs. Importantly, market prices reveal information about wealth-weighted beliefs $\mathbb{P}_{\text{imp}}$ but it is  unweighted beliefs, $\mathbb{P}$, which generate the public signal. 

One is led to the question: Why do people continue to bet with each other when these bets are highly risky? The answer we propose is that everyone in our economy thinks that the market is wrong and that by betting, they will be able to make money on average. They do not use the implied probability revealed by the markets to improve their estimate of $\mathbb{P}$, since this trading strategy would be (subjectively) sub-optimal. Quite remarkably, the coupled dynamics of individual wealth and beliefs leads to a fat-tailed distribution of wealth. The richest agents at a given instant in time are not necessarily the smartest ones but rather those who have made bold, successful bets in the past. Since those agents tend to dominate the market, the implied probability $\mathbb{P}_{\text{imp}}$ cannot be used to learn the true probability $\mathbb{P}$.\footnote{Introducing a wealth tax, or an inheritance tax, tends to reduce inequalities and, in our model, helps prices reveal private beliefs -- see Appendix E.}   

Why are there no Warren Buffets who invest for the very long run by guessing that the probability of a successful outcome will be equal to the mean $\mathbb{P}=1/2$ of the invariant distribution? Our answer is that this would only be the case if we lived forever and could afford to be strict Bayesian learners, but the world that we live in is far better approximated by observing recent realizations than by relying on an unconditional long-run ergodic measure. We believe that our quasi-non-ergodic model aptly illustrates what Keynes had in mind when he wrote that ``In the long run we are all dead''. 

\bibliography{FJPlib.bib}

\begin{thebibliography}{55}
\providecommand{\natexlab}[1]{#1}
\providecommand{\url}[1]{\texttt{#1}}
\expandafter\ifx\csname urlstyle\endcsname\relax
  \providecommand{\doi}[1]{doi: #1}\else
  \providecommand{\doi}{doi: \begingroup \urlstyle{rm}\Url}\fi

\bibitem[Adam et~al.(2016)Adam, Marcet, and Nicolini]{ada-mar-nic-1}
Klaus Adam, Albert Marcet, and Juan~Pablo Nicolini.
\newblock Stock market volatility and learning.
\newblock \emph{Journal of Finance}, 71\penalty0 (1):\penalty0 33--82, 2016.

\bibitem[Adam et~al.(2017)Adam, Marcet, and Beutel]{ada-mar-beu-1}
Klaus Adam, Albert Marcet, and Johannes Beutel.
\newblock Stock price booms and expected capital gains.
\newblock \emph{American Economic Review}, 107\penalty0 (8):\penalty0
  2352--2408, 2017.

\bibitem[Alchian(1970)]{AlchianInfo}
Armen~A. Alchian.
\newblock Information costs, pricing, and resource unemployment.
\newblock In Edmund~S. Phelps, G.~C. Archibald, and Armen~A. Alchian, editors,
  \emph{Microeconomic Foundations of Employment and Inflation Theory}. Norton,
  New York, 1970.

\bibitem[Anderson(1989)]{Anderson_Spin6}
Philip~W. Anderson.
\newblock Spin glass vi: Spin glass as cornucopia.
\newblock \emph{Physics Today}, 42\penalty0 (9):\penalty0 9--11, 1989.

\bibitem[Angeletos and {La'O}(2011)]{ang-lao-1}
George-Marios Angeletos and {La'O}.
\newblock Sentiments.
\newblock \emph{Econometrica}, 81\penalty0 (2):\penalty0 739--779, 2011.

\bibitem[Beker and Espino(2011)]{bek-esp-1}
Pablo Beker and Emilio Espino.
\newblock The dynamics of efficient asset trading with heterogeneous beliefs.
\newblock \emph{Journal of Economic Theory}, 166:\penalty0 189--229, 2011.

\bibitem[Benhabib and Bisin(2018)]{ben-bis-1}
Jess Benhabib and Alberto Bisin.
\newblock Skewed wealth distributions: Theory and empirics.
\newblock \emph{Journal of Economic Literature}, 56\penalty0 (4):\penalty0
  1261--1291, 2018.

\bibitem[Benhabib and Chetan(2014)]{ben-che-1}
Jess Benhabib and Dave Chetan.
\newblock Learning, large deviations and rare events.
\newblock \emph{Review of Economic Dynamics}, 17\penalty0 (3):\penalty0
  367--382, 2014.

\bibitem[Benhabib et~al.(2011)Benhabib, Bisin, and Zhu]{ben-bis-zhu-1}
Jess Benhabib, Alberto Bisin, and Shenghao Zhu.
\newblock The distribution of wealth and fiscal policy in economies with
  finitely lived agents.
\newblock \emph{Econometrica}, 79\penalty0 (1):\penalty0 123--157, 2011.

\bibitem[Benhabib et~al.(2015)Benhabib, Wang, and Wen]{ben-wan-wen-1}
Jess Benhabib, Pengfei Wang, and Yi~Wen.
\newblock Sentiments and aggregate demand fluctuations.
\newblock \emph{Econometrica}, 83\penalty0 (2):\penalty0 549--585, 2015.

\bibitem[Blanchard(1985)]{BlanchardFiniteHorizons}
Olivier~J. Blanchard.
\newblock Debt, deficits, and finite horizons.
\newblock \emph{Journal of Political Economy}, 93\penalty0 (April):\penalty0
  223--247, 1985.

\bibitem[Blume and Easley(2006)]{blu-eas-1}
Lawrence Blume and David Easley.
\newblock If you’re so smart, why aren’t you rich?
\newblock \emph{Econometrica}, 74\penalty0 (4):\penalty0 929--966, 2006.

\bibitem[Borovi\v{c}ka(2000)]{jar-bor-1}
Jarislov Borovi\v{c}ka.
\newblock Survival and long-run dynamics with heterogeneous beliefs under
  recursive preferences.
\newblock \emph{Journal of Political Economy}, 128\penalty0 (1):\penalty0
  206--251, 2000.

\bibitem[Bouchaud(2013)]{Bouchaud2013}
Jean-Philippe Bouchaud.
\newblock Crises and collective socio-economic phenomena: simple models and
  challenges.
\newblock \emph{Journal of Statistical Physics}, 151\penalty0 (3-4):\penalty0
  567--606, 2013.

\bibitem[Bouchaud(2019)]{Bouchaud2021}
Jean-Philippe Bouchaud.
\newblock Radical complexity.
\newblock \emph{Royal Economic Society}, 2019.
\newblock URL \url{https:// www.res.org.uk/ resources-page/
  radical-complexity.html}.

\bibitem[Bouchaud and M\'ezard(2000)]{bouchaud_mezard}
Jean-Philippe Bouchaud and Marc M\'ezard.
\newblock Wealth condensation in a simple model of the economy.
\newblock \emph{Physica A: Statistical Mechanics and its Applications},
  282\penalty0 (3-4):\penalty0 536--545, 2000.

\bibitem[Boyd et~al.(2004)Boyd, Diaconis, and Xiao]{Diaconis}
Stephen Boyd, Persi Diaconis, and Lin Xiao.
\newblock Fastest mixing markov chain on a graph.
\newblock \emph{SIAM Review}, 46\penalty0 (4):\penalty0 667--689, 2004.
\newblock \doi{10.1137/S0036144503423264}.
\newblock URL \url{https://doi.org/10.1137/S0036144503423264}.

\bibitem[Brock and Durlauf(2001)]{brock}
William~A. Brock and Steven~N. Durlauf.
\newblock {Discrete Choice with Social Interactions}.
\newblock \emph{The Review of Economic Studies}, 68\penalty0 (2):\penalty0
  235--260, 04 2001.

\bibitem[Cass and Shell(1983)]{cassShellSunspots}
Dave Cass and Karl Shell.
\newblock Do sunspots matter?
\newblock \emph{Journal of Political Economy}, 91:\penalty0 193--227, 1983.

\bibitem[Cogley and Sargent(2008)]{cog-sar-1}
Timothy Cogley and Thomas~J. Sargent.
\newblock The market price of risk and the equity premium: A legacy of the
  great depression?
\newblock \emph{Journal of Monetary Economics}, 5\penalty0 (3):\penalty0
  454--476, April 2008.

\bibitem[Cogley and Sargent(2009)]{cog-sar-2}
Timothy Cogley and Thomas~J. Sargent.
\newblock Diverse beliefs, survival and the market price of risk.
\newblock \emph{Economic Journal}, 119:\penalty0 354--376, March 2009.

\bibitem[Daniel~Kahneman(2021)]{Kahneman2021}
Cass R.~Sunstein Daniel~Kahneman, Olivier~Sibony.
\newblock \emph{Noise: A Flaw in Human Judgment}.
\newblock Little, Brown Spark, New York, first edition, 2021.

\bibitem[Debenedetti and Stillinger(2001)]{Stillinger}
Pablo~G. Debenedetti and Frank~H. Stillinger.
\newblock Supercooled liquids and the glass transition.
\newblock \emph{Nature}, 410\penalty0 (6825):\penalty0 259--267, 2001.
\newblock URL \url{http://dx.doi.org/10.1038/35065704}.

\bibitem[Evans and Honkapohja(2001)]{EvansHonkLeanringBook}
George~W. Evans and Seppo Honkapohja.
\newblock \emph{Learning and Expectations in Macroeconomics}.
\newblock Princeton University Press, Princeton, 2001.

\bibitem[Evans and Honkapohja(2013)]{EvansHonk2013}
George~W. Evans and Seppo Honkapohja.
\newblock Learning as a rational foundation for macroeconomics and finance.
\newblock In Roman Frydman and Edward~S. Phelps, editors, \emph{Rethinking
  Expectations: The Way Forward for Macroeconomics}, pages 68 -- 111. Princeton
  University Press, Princeton, 2013.

\bibitem[Farmer(1999)]{FarmerProphesiesBook1999}
Roger E.~A. Farmer.
\newblock \emph{The Macroeconomics of Self-Fulfilling Prophecies}.
\newblock MIT Press, Cambridge, MA, second edition, 1999.

\bibitem[Farmer(2018)]{Far-ass-perpet}
Roger E.~A. Farmer.
\newblock Pricing assets in a perpetual youth model.
\newblock \emph{Review of Economic Dynamics}, 30:\penalty0 106--124, 2018.

\bibitem[Farmer(2020)]{far-ind-sch}
Roger E.~A. Farmer.
\newblock The indeterminacy school in macroeconomics.
\newblock \emph{Oxford Research Encyclopedia of Economics and Finance}, online
  publication, April 2020.

\bibitem[Farmer(2021)]{far-imp-bel}
Roger E.~A. Farmer.
\newblock The importance of beliefs in shaping macroeconomic outcomes.
\newblock \emph{Oxford Review of Economic Policy}, forthcoming, 2021.

\bibitem[Farmer et~al.(2011)Farmer, Nourry, and Venditti]{far-nou-ven-2}
Roger E.~A. Farmer, Carine Nourry, and Alain Venditti.
\newblock Debt deficits and finite horizons, the stochastic case.
\newblock \emph{Economics Letters}, 111:\penalty0 47--49, 2011.

\bibitem[Friedman(1953)]{FriedmanPositive}
Milton Friedman.
\newblock \emph{Essays in Positive Economics}.
\newblock University of Chicago Press, Chicago, 1953.

\bibitem[Gabaix(2009)]{gabaix_tails}
Xavier Gabaix.
\newblock Power laws in economics and finance.
\newblock \emph{Annual Review of Economics}, 1\penalty0 (1):\penalty0 255--294,
  2009.

\bibitem[Gabaix et~al.(2016)Gabaix, Lasry, Lyons, and Moll]{gabaix}
Xavier Gabaix, Jean-Michel Lasry, Pierre-Louis Lyons, and Benjamin Moll.
\newblock The dynamics of inequality.
\newblock \emph{Econometrica}, 84\penalty0 (6):\penalty0 2071--2111, 2016.

\bibitem[Galla and Farmer(2013)]{galla-farmer}
Tobias Galla and J.~Doyne Farmer.
\newblock Complex dynamics in learning complicated games.
\newblock \emph{Proceedings of the National Academy of Sciences}, 110\penalty0
  (4):\penalty0 1232--1236, 2013.
\newblock ISSN 0027-8424.
\newblock \doi{10.1073/pnas.1109672110}.
\newblock URL \url{https://www.pnas.org/content/110/4/1232}.

\bibitem[G\^arleanu and Panageas(2015)]{Garleanu}
Nicolae G\^arleanu and Stavros Panageas.
\newblock Young, old, conservative, and bold: The implications of heterogeneity
  and finite lives for asset pricing.
\newblock \emph{Journal of Political Economy}, 123\penalty0 (3):\penalty0
  670--685, 2015.
\newblock \doi{10.1086/680996}.
\newblock URL \url{https://doi.org/10.1086/680996}.

\bibitem[Grossman and Stiglitz(1980)]{gro-sti-1}
Sanford Grossman and Joseph~E. Stiglitz.
\newblock On the impossibility of informationally efficient markets.
\newblock \emph{American Economic Review}, 70\penalty0 (3):\penalty0 393--408,
  1980.

\bibitem[H\'anggi et~al.(1990)H\'anggi, Talkner, and Borkovec]{hanggi}
Peter H\'anggi, Peter Talkner, and Michal Borkovec.
\newblock Reaction-rate theory: fifty years after {K}ramers.
\newblock \emph{Reviews of modern physics}, 62:\penalty0 251--341, 1990.

\bibitem[Horst(2017)]{horst}
Ulrich Horst.
\newblock \emph{Ergodicity and Nonergodicity in Economics}, pages 1--6.
\newblock Palgrave Macmillan UK, London, 2017.

\bibitem[{Kelly Jr.}(1956)]{joh-kel-1}
John~Larry {Kelly Jr.}
\newblock A new interpretation of information rate.
\newblock \emph{Bell System Technical Journal}, 35\penalty0 (4):\penalty0
  917--926, 1956.

\bibitem[Kesten(1973)]{kesten}
Harry Kesten.
\newblock Random difference equations and renewal theory for products of random
  matrices.
\newblock \emph{Acta Math.}, 131:\penalty0 207--248, 1973.
\newblock \doi{10.1007/BF02392040}.
\newblock URL \url{https://doi.org/10.1007/BF02392040}.

\bibitem[Keynes(1936)]{KeynesGeneralTheory}
John~Maynard Keynes.
\newblock \emph{The General Theory of Employment, Interest and Money}.
\newblock MacMillan and Co., London and Basingstoke, 1936.
\newblock 1973 edition published for the Royal Economic Society, Cambridge.

\bibitem[Kirman(1993)]{kirman}
Alan Kirman.
\newblock Ants, rationality, and recruitment.
\newblock \emph{The Quarterly Journal of Economics}, 108:\penalty0 137--156,
  1993.

\bibitem[{Lucas Jr.}(1972)]{LucasExpectationsNeutrality}
Robert~E. {Lucas Jr.}
\newblock Expectations and the neutrality of money.
\newblock \emph{Journal of Economic Theory}, 4:\penalty0 103--124, 1972.

\bibitem[Massari(2019)]{Massari}
Filippo Massari.
\newblock Market selection in large economies: A matter of luck.
\newblock \emph{Theoretical Economics}, 14\penalty0 (2):\penalty0 437--473,
  2019.
\newblock \doi{https://doi.org/10.3982/TE2456}.
\newblock URL \url{https://onlinelibrary.wiley.com/doi/abs/10.3982/TE2456}.

\bibitem[Moran et~al.(2020{\natexlab{a}})Moran, Fosset, Benzaquen, and
  Bouchaud]{Moran_2020}
Jos{\'{e}} Moran, Antoine Fosset, Michael Benzaquen, and Jean-Philippe
  Bouchaud.
\newblock Schrodinger's ants: a continuous description of kirman's recruitment
  model.
\newblock \emph{Journal of Physics: Complexity}, 1\penalty0 (3):\penalty0
  035002, Aug 2020{\natexlab{a}}.

\bibitem[Moran et~al.(2020{\natexlab{b}})Moran, Fosset, Luzzati, Bouchaud, and
  Benzaquen]{trapping}
Jos\'e Moran, Antoine Fosset, Davide Luzzati, Jean-Philippe Bouchaud, and
  Michael Benzaquen.
\newblock By force of habit: Self-trapping in a dynamical utility landscape.
\newblock \emph{Chaos: An Interdisciplinary Journal of Nonlinear Science},
  30\penalty0 (5):\penalty0 053123, 2020{\natexlab{b}}.
\newblock \doi{10.1063/5.0009518}.

\bibitem[Moran(1958)]{moran}
Patrick~A.P. Moran.
\newblock Random processes in genetics.
\newblock \emph{Mathematical Proceedings of the Cambridge Philosophical
  Society}, 54:\penalty0 60--71, 1958.

\bibitem[Morelli et~al.(2020)Morelli, Benzaquen, Tarzia, and Bouchaud]{morelli}
Federico~Guglielmo Morelli, Michael Benzaquen, Marco Tarzia, and Jean-Philippe
  Bouchaud.
\newblock Confidence collapse in a multihousehold, self-reflexive dsge model.
\newblock \emph{Proceedings of the National Academy of Sciences}, 117\penalty0
  (17):\penalty0 9244--9249, 2020.
\newblock ISSN 0027-8424.
\newblock \doi{10.1073/pnas.1912280117}.
\newblock URL \url{https://www.pnas.org/content/117/17/9244}.

\bibitem[Morris and Shin(2002)]{mor-shi-1}
Stephen Morris and Hyun~Song Shin.
\newblock Social value of public information.
\newblock \emph{The American Economic Review}, 92\penalty0 (5):\penalty0
  1521--1534, 2002.

\bibitem[Parisi(2007)]{Parisi2007}
G.~Parisi.
\newblock Physics, complexity and biology.
\newblock \emph{Advances in Complex Systems}, 10\penalty0 (supp02):\penalty0
  223--232, 2007.
\newblock \doi{10.1142/S021952590700132X}.
\newblock URL \url{https://doi.org/10.1142/S021952590700132X}.

\bibitem[Pemantle(2007)]{pemantle}
Robin Pemantle.
\newblock A survey of random processes with reinforcement.
\newblock \emph{Probability surveys}, 4:\penalty0 1--79, 2007.

\bibitem[Peters(2019)]{peters}
Ole Peters.
\newblock The ergodicity problem in economics.
\newblock \emph{Nature Physics}, 15\penalty0 (12):\penalty0 1261--1221, 2019.

\bibitem[Sandroni(2000)]{alv-san-1}
Alvaro Sandroni.
\newblock Do markets favor agents able to make accurate predictions?
\newblock \emph{Econometrica}, 68\penalty0 (6):\penalty0 1303--1342, 2000.

\bibitem[Stokey et~al.(1989)Stokey, Lucas, and
  Prescott]{StokeyLucasPrescottBook}
Nancy~L. Stokey, Robert~E. Lucas, Jr., and with Edward~C. Prescott.
\newblock \emph{Recursive Methods in Economic Dynamics}.
\newblock Harvard University Press, Cambridge, MA, 1989.

\bibitem[Young(2002)]{P_Young}
H.~P. Young.
\newblock \emph{Individual strategy and social structure}.
\newblock Princeton University Press, 2002.

\end{thebibliography}

\appendix 
\newpage

\section{Appendix A: The continuous time limit}\label{newapp}

\subsection{Derivation of Eq. \eqref{eq_Pbeta}}


Introducing a change of variable $u$ such that $\mathbb{P}=\frac12 + u$, one can convert Eq. \eqref{eq_master} into:
\begin{multline}\label{FPu}
(1-\hat \delta)^2 \mathcal{P}_{t+1}(u) = \frac{1 - \hat \delta - \hat \lambda}{2} \left[\mathcal{P}_{t}\left(\frac{u-\hat \lambda/2}{1 - \hat \delta}\right) + \mathcal{P}_{t}\left(\frac{u+\hat \lambda/2}{1 - \hat \delta}\right)\right] \\  + u \left[\mathcal{P}_{t}\left(\frac{u-\hat \lambda/2}{1 - \hat \delta}\right) - \mathcal{P}_{t}\left(\frac{u+\hat \lambda/2}{1 - \hat \delta}\right)\right]
\end{multline}
where $\hat \lambda := \lambda ( 1 - \delta)$ and $\hat \delta := \delta + \hat \lambda$. Note that this equation preserves the symmetry $\mathcal{P}_{t}(-u)=\mathcal{P}_{t}(u)$ (i.e. $\mathbb{P} \to 1 - \mathbb{P}$) valid for all times.

In the following analysis we  assume long memory ($\lambda \ll 1$) and long lifetimes ($\delta \ll 1$) by  focusing on the limit where $\lambda, \delta \to 0$ with $\delta = \alpha \lambda^2$ for fixed $\alpha=O(1)$. Expanding Eq. \eqref{FPu} to order $\lambda^3$ yields:
\begin{multline} \label{FP_discrete}
\Delta_t = \delta \left[ u \mathcal{Q} \right]' + \frac{\lambda^2}{2} \left[(\frac14 - u^2)\mathcal{Q}\right]'' - 2 \lambda \delta \left[u^2 \mathcal{Q}\right]''  \\ - \frac{\lambda^3}{2}  \left[(\frac{u}{12}-\frac{u^3}{3})\mathcal{Q}'' - u^2 \mathcal{Q}' + \frac{5}{12} \mathcal{Q}' 
\right]'+ O(\lambda^4),
\end{multline}
where primes denote derivatives with respect to $u$, $\mathcal{P}(u)\equiv (1- \hat \delta)\mathcal{Q}(u (1- \hat \delta))$, and  $\Delta_t\equiv \mathcal{Q}_{t+1}(u)-\mathcal{Q}_{t}(u)$. Note that the last two terms of Eq.\ \eqref{FP_discrete}  are of order $\lambda^3$, and we will neglect them in the following approximation. 

In the small $\delta, \lambda$ limit, Eq.\ \eqref{FP_discrete} converges to  the following continuous time Fokker-Planck equation for $\mathcal{P}$:
\be \label{FP_cont}
\frac{1}{\lambda^2} \frac{\partial \mathcal{P}}{\partial t} =  \alpha \left[ u \mathcal{P} \right]' + \frac{1}{2} \left[(\frac14 - u^2)\mathcal{P}\right]''.
\ee
This equation coincides with the continuous time description of Kirman's ant recruitment model \citep{kirman}, for which a lot is known (see  \citet{Moran_2020} for recent results and references).

In particular the stationary distribution $\mathcal{P}^*$ is is described by the following second order differential equation.
\begin{equation}
\alpha \left[ u \mathcal{P}^* \right]' + \frac{1}{2} \left[(\frac14 - u^2)\mathcal{P}^*\right]'' = 0.
\end{equation}
The solution to this equation is given by
\begin{equation}\label{eq_Pbeta1}
\mathcal{P}_\infty(u) = \frac{\Gamma(2\alpha)}{\Gamma^2(\alpha)}\left(\frac14 - u^2\right)^{\alpha-1},
\end{equation}
which corresponds to Eq.\ \eqref{eq_Pbeta} in the text.

\subsection{Generalization: non-linear feedback}
\label{sigmoid} 

The Fokker-Planck equation Eq. \eqref{FP_cont} corresponds to the following stochastic differential equation:
\be 
{\rm d}\mathbb{P} = -\delta (\mathbb{P} - \frac12) {\rm d}t + \lambda \sqrt{\mathbb{P}(1-\mathbb{P})} {\rm d}W_t,
\ee
where $W_t$ is a Wiener noise. More generally, one can consider a sigmoidal feedback term $\mathcal{F}(\mathbb{P})$ mapping the average belief onto the true probability,
\be
\mathbb{P}_{t+1} = \mathcal{F}(\mathbb{P}_t)
\ee 
with $\mathcal{F}(\mathbb{P})=\mathbb{P}$ throughout the main part of the paper and in section above. In this case, one obtains as a stochastic differential equation
\be 
{\rm d}\mathbb{P} = - \partial_\mathbb{P} \mathcal{V}(\mathbb{P}) {\rm d}t + \lambda \sqrt{\mathbb{P}(1-\mathbb{P})} {\rm d}W_t,
\ee
where we have introduced a ``potential function'' $\mathcal{V}(x)$ such that
\be
\partial_x \mathcal{V}(x) := \delta(x-\frac12) + \lambda(x-\mathcal{F}(x)).
\ee
For definiteness, consider a sigmoidal function $\mathcal{F}(x)$ defined as:
\be
\mathcal{F}(x) = \frac12 \left(1 + \tan [\zeta(x-\frac12)] \right)
\ee 
The corresponding potential $\mathcal{V}(x)$ is then given by
\be
\mathcal{V}(x) = \frac12 (\delta+ \lambda) u^2 - \frac{\lambda}{2 \zeta} \log \cosh \zeta u; \qquad u:=x-\frac12
\ee
For small $\zeta$, $\mathcal{V}(x)$ has a unique minimum corresponding to $x=1/2$. For $\zeta > \zeta_c=2(1+\delta/\lambda)$, $\mathcal{V}(x)$ has two minima $x^* < 1/2$ and $1-x^*> 1/2$ and one maximum at $x=1/2$.  

In the absence of the Wiener noise term, the dynamics of $x$ would just be ``rolling down'' the potential slopes, selecting one of the minima of $\mathcal{V}(x)$ (corresponding to the stable solutions of $\mathcal{F}(x)=x$). 

In the presence of noise and for $\zeta > \zeta_c$, the dynamics becomes a succession of long phases where $\mathbb{P}_{t}$ remains close to either $x^*$ or $1-x^*$, separated by rapid switches from one minimum to the other. The time $\tau_\times$ needed to ``climb up the hill'' separating the two minima can be however very long when $\lambda \to 0$. 

In fact, this time can be rather accurately computed by changing variables from $\mathbb{P}$ to 
$\phi$ where $\mathbb{P}=(1+\sin \phi)/2$, which allows one to get rid of the factor $\sqrt{\mathbb{P}(1-\mathbb{P})}$ in front of the Wiener noise, see e.g. \cite{Moran_2020}. Using a standard approach (e.g. \cite{hanggi}), one can then show that
\[
\tau_\times \sim \lambda^{-1} e^{\Gamma/\lambda}, \qquad (\lambda \to 0),
\]
where $\Gamma$ can be fully computed (at least numerically) for {\it any} potential $\mathcal{V}(x)$. The exponential dependence of $\tau_\times$ in $\lambda$ implies that (a) there is a strong separation of timescales in such models and (b) the precise value of $\tau_\times$ is unknowable in practice, as it is highly sensitive on the detailed value of the parameters of the model. Hence agents cannot be assumed to use the same learning rule. Since these switches can be interpreted as ``crashes'', the probability of such crashes is, in our simple model, unknowable much as the trajectories of a chaotic system are unknowable (for a related discussion, see \cite{morelli}). 

\section{Appendix B: Dispersion of opinions}
\label{app_dispersion}
Taking the expectation of Eq. \eqref{eq_D} over the realization of $s_t$ one gets:
\be 
\mathbb{E}[\mathbb{D}_{i,t+1}] = 
(1-\delta)\left[(1-\lambda)\mathbb{E}[\mathbb{D}_{i,t}]+\delta(\mathbb{P}_t -\frac12)\right]+\delta (1-\delta)\left[\frac12 - \mathbb{P}_t\right],
\ee 
or
\be 
\mathbb{E}[\mathbb{D}_{i,t+1}] = 
(1-\delta)(1-\lambda)\mathbb{E}[\mathbb{D}_{i,t}]
\ee 
which shows that $\mathbb{E}[\mathbb{D}_{i,t}]$ tends to zero when $t \to \infty$. 

Now let us square Eq. \eqref{eq_D} before taking the average over $s_t$. One now gets:
\begin{align}
\mathbb{E}[\mathbb{D}_{i,t+1}^2] &=
(1 - \delta)\left[(1-\lambda)^2 
\mathbb{E}[\mathbb{D}_{i,t}^2] + \delta^2 \mathbb{E}[(\mathbb{P}_t-\frac12)^2]\right] \\ \nonumber 
&+\delta\left[\mathbb{E}[z^2]+\frac{\delta^2}{4}-\frac{\delta}{2}+(1-\delta)^2(1-\lambda^2)(\mathbb{P}_t^2-\mathbb{P}_t)\right].
\end{align}
Now taking further the expectation over the distribution $\mathcal{P}$ of the probability $\mathbb{P}$, and using
\be 
\mathbb{E}_{\mathcal{P}}[\mathbb{P}^2] = \frac{1+\alpha}{2(1+2 \alpha)}, \qquad \alpha = \frac{\delta}{\lambda^2},
\ee 
we obtain, in the limit $\delta, \lambda \to 0$, with $\alpha$ fixed,
\be 
\mathbb{E}^\star[\mathbb{D}_{i,t+1}^2] =
(1-\delta)(1-\lambda)^2 
\mathbb{E}^\star[\mathbb{D}_{i,t}^2] + \frac{\delta}{6} \, \frac{2+\alpha}{1 + 2 \alpha} + O(\delta^2),
\ee 
where $\mathbb{E}^\star$ means an expectation both over $s$ and $\mathcal{P}$.

Hence in the stationary state where $\mathbb{E}^\star[\mathbb{D}_{i,t}^2]$ is independent of $t$ one finds:
\be 
\mathbb{E}^\star[\mathbb{D}_{i}^2] \approx \frac{\delta}{6(1 - (1-\delta)(1-\lambda)^2)} \, \frac{2+\alpha}{1 + 2 \alpha},
\ee 
and hence the result Eq. \eqref{eq_dispersion}.

\section{Appendix C: Solving the individual optimization problem} \label{appsolve}
We conjecture that the value function has the form
\begin{equation}
    A \log W_i(\sigma) + B,
\end{equation}
for unknown constants $A$ and $B$.
Substituting from Eq.\ \eqref{E3} for $c_i(\sigma)$ in Eq.\ \eqref{E1} and taking derivatives with respect to $W_i(\sigma')$ leads to the following  Euler equation,
\begin{equation} \label{B2}
 \frac{x_i(\sigma') Q(\sigma'|\sigma )}{c_i(\sigma)}   = \frac{A \beta  \mathbb{P}_i(\sigma')x_i(\sigma')  }{W_i(\sigma')},
\end{equation}
which holds state by state. Using the envelope condition $ A c_i(\sigma) = W_i(\sigma),$  which holds at every date and in every state, we can write Eq.\ \eqref{B2} as
\begin{equation}\label{B4}
        x_i(\sigma') Q(\sigma'|\sigma )W_i(\sigma') =  \beta  \mathbb{P}_i(\sigma')x_i(\sigma')  W_i(\sigma).
\end{equation}
Combining the budget equation, Eq.\ \eqref{E3}, which holds with equality with Eq.\ \eqref{B4} leads to the expression,
\be \label{E3a}
\sum_{\sigma'}  \beta  \mathbb{P}_i(\sigma')x_i(\sigma')  W_i(\sigma) + \frac{W_i(\sigma)}{A} =  W_i(\sigma) .
\ee
Because $s'$ is independent of $\xx'$ 
\be
\sum_{\sigma'} \mathbb{P}_i(\sigma')x_i(\xx') =\sum_{\xx'}p(\xx') x_i(\xx') \sum_{s'} \mathbb{P}_i(s') = 1-\delta
\ee
and thus by canceling terms and rearranging Eq.\ \eqref{B4} we arrive at the following value for $A.$
\be \label{E3aa}
A = \frac{1}{1-\beta (1-\delta)}
\ee
The constant $B$ does not affect the solution and can be solved for by plugging the value of $A$ into the expression
\be 
A \log(W_i) + B  = \log \left(\frac{W_i}{A}\right) + \beta (1-\delta) \left[A \log(W_i) +B \right]
\ee 
and equating the coefficients on the constant terms.

It follows from Eq.\ \eqref{B4} that for all $x_i(\xx')=1$, that is, those who survive,
\be \label{E31}
W_i(\sigma') = \beta \frac{ \mathbb{P}_i(\sigma')}{Q(\sigma' | \sigma)}  W_i(\sigma).
\ee
This establishes the first term on the right side of Eq.\ \eqref{E6}. If $x_i(\xx')=0$ the newborn with index $i$ has wealth $H$ by assumption. This establishes the second term on the right side of  Eq. \eqref{E6}.

\section{Appendix D: Establishing the Properties of Equilibrium} \label{appsolve1}
From Eq.\ \eqref{HWEQ}, we have the following equation for human wealth,
\be \label{stuff}
H_i(\sigma) = \varepsilon +     \sum_{\sigma'} Q(\sigma'|\sigma) \, x_i' \,  H_i(\sigma'). 
\ee
From the definition of total wealth we have that $W_i(\sigma') - H_i(\sigma')  = a_i(\sigma')$ where $a_i(\sigma')$ is the amount of Arrow security held by agent $i$ that pays one unit if $\sigma'$ is realized. Assuming market clearing means that for each $\sigma'$, 
\be \label{a0}
\sum_{i=1}^N a_i(\sigma') = 0, \qquad \forall \sigma',
\ee 
and hence, using Eq.\ \eqref{E31},  we have that
\be \label{D3}
\sum_{i=1}^N W_i(\sigma') = N(\sigma') H_i(\sigma')  = \beta \frac{1}{Q(\sigma' | \sigma)}  \sum_{i=1}^N \mathbb{P}_i(\sigma') W_i(\sigma).
\ee 
Rearranging this equation and factoring $\mathbb{P}_i(\sigma')$ as
$p(\xx')\mathbb{P}_i(s')$ gives the following expression for the pricing kernel
\be \label{E31bis}
 Q(\sigma'|\sigma) = \beta  p(\xx') \frac{\sum_{i=1}^N  \mathbb{P}_i(s') x_i' W_i(\sigma)}{N(\sigma') H_i(\sigma')},
\ee
which establishes Eq.\ \eqref{Qeqb} from Proposition \ref{Prop2}.

Replacing Eq.\ \eqref{E31bis} in Eq.\ \eqref{stuff}  and reversing the order of summation gives
\be \label{stuff1}
H(\sigma) = \varepsilon +    \sum_{i=1}^N W_i(\sigma) \sum_{\sigma'}\left\{  \frac{\beta}{N(\sigma') H(\sigma')}  \mathbb{P}_i(s') p(\xx') \, x_i' \,  H(\sigma') \right\}. 
\ee
Next, cancel $H(\sigma')$ from top and bottom,
 \be 
H(\sigma) = \varepsilon +    \beta \sum_{i=1}^N  W_i(\sigma) \sum_{\xx'}\left\{  \frac{\beta p(\xx')  \, x_i'}{N(\xx') }   \,   \right\}  \sum_{s'} \mathbb{P}_i(s'). 
\ee
Using the facts that $\mathbb{P}_i(s')=1$,  $\sum_{\xx'}\left\{  \frac{ p(\xx')  \, x_i'}{N(\xx') }   \,   \right\} =1-\delta$ and $\sum_{i=1}^N W_i(\sigma)= H(\sigma)  $ this expression simplifies to,
\be 
H(\sigma) =  \varepsilon +    \beta    H(\sigma) (1-\delta), 
\ee
or
\be 
H(\sigma) = \frac{ \varepsilon} {1-\beta(1-\delta)}
\ee
which established Eq.\ \eqref{HW} in Proposition \ref{Prop2}.

\section{Appendix E: Proof of Proposition \ref{prop3}}\label{proof3}
We now seek an expression for the price of a security that pays a dividend $d$ every time  $s_t=\{1\}$. This is given by the expression,
\be \label{F1}
p_{\text{E}}(\sigma) = \sum_{\sigma'} Q(\sigma'|\sigma) \left[ d \, \delta_{s',1} + p_{\text{E}}'(\sigma') \right]
\ee
where $\sigma' = (\xx',s')$ is tomorrow's state, with $\xx'$ encoding who survives and who dies and $\delta_{s',1}$ is the index function which equals $1$ when $s'=1$ and 0 otherwise. Iterating Eq.\ \eqref{F1}  gives the following infinite series:
\be \label{F2}
p_{\text{E}}(\sigma) = d  \sum_{\sigma'} Q(\sigma'|\sigma)  \delta_{s',1} + d  \sum_{\sigma',\sigma''} Q(\sigma'|\sigma) Q(\sigma''|\sigma') \delta_{s'',1} + \cdots,
\ee
where, from Eq. \eqref{eq_full_implied},
\be
   Q(\sigma'|\sigma) = \beta p(\xx')  \left(  \frac{ \sum_{i=1}^N\mathbb{P}_i(s') W_i(\sigma) x'_i} {N(\sigma') H } \right).
\ee
As we have shown in the main text, this object converges, for large $N$, to
\be 
Q(\sigma'|\sigma) = \beta p(\xx') \mathbb{P}_{\text{imp}}(s'),
\ee where
\[ 
\mathbb{P}_{\text{imp}}(s') := \frac{1}{NH} \sum_{i=1}^N\mathbb{P}_i(s') W_i(s).
\] 
Hence,
\be 
\sum_{\sigma'} Q(\sigma'|\sigma)  \delta_{s',1} \equiv \beta \mathbb{P}_{\text{imp}}
\ee 
where recall that dropping the argument $s$ implicitly means $s = \{1\}$. The first contribution to $p_\text{E}$ is thus simply
\[ 
d \beta \mathbb{P}_{\text{imp}}.
\]

Now let us turn to the second term, which takes the form
\begin{multline}
\sum_{\sigma'} Q(\sigma''|\sigma') Q(\sigma'|\sigma)   \\ =
\frac{\beta p(\xx'')}{N(\sigma'')H} \sum_{\sigma'} \sum_j x_j(\xx'')  \mathbb{P}'_j(s''|s') W_j'(s') \, Q(\sigma'|\sigma). 
\end{multline}
Expressing $W_j'(s')$ thanks to Eq. \eqref{eq_full_implied_2}, the right-hand side reads:
\begin{multline}\label{F7}
\frac{\beta}{N(\sigma'')H} \left[ \sum_{j,\sigma'} \beta x_j(\xx'') p(\xx'') \mathbb{P}'_j(s''|s') x_j(\xx') \mathbb{P}_j(s') W_j(s) \right. \\ + \left. \sum_{j,s'} x_j(\xx'') p(\xx'') \mathbb{P}'_j(s''|s') (1 - x_j(\xx')) H Q(\sigma'|\sigma) \right],
\end{multline} 
where the first term corresponds to surviving agents in the next time step, and the second term to dying agents that are replaced with new born agents with wealth $H$.

Consider the two terms of Eq.\ \eqref{F7} in turn. The first term contains a factor $x_j(\xx'')x_j(\xx')$ which equals 1 if  an  agent $j$  survives for both of the next two periods and zero otherwise. We now use the update rule of agents' beliefs to compute $\mathbb{P}'_j(\sigma''|\sigma')$. One finds, for $s''=\{1\}$,
\[
\mathbb{P}'_j(1|1) = (1 - \lambda) \mathbb{P}_j + \lambda; \qquad  \mathbb{P}'_j(1|0) = (1 - \lambda) \mathbb{P}_j,
\]
where we recall that $\mathbb{P}_j :=\mathbb{P}_j(1)$. Hence
\[
\sum_{s'} \mathbb{P}'_j(1|s') \mathbb{P}_j(s') = \left[(1 - \lambda) \mathbb{P}_j + \lambda\right] \mathbb{P}_j + \left[(1 - \lambda) \mathbb{P}_j\right](1 - \mathbb{P}_j)=
\mathbb{P}_j.
\]
In words, conditional on survival, the agent's belief is a martingale. Conditioning on  $s''=\{1\}$, one has:
\[
\sum_{\xx'',s''=\{1\}} \beta  p(\xx'') \sum_{j,\sigma'} x_j(\xx'') \mathbb{P}'_j(s''|s') x_j(\xx') p(\xx') \mathbb{P}_j(s') W_j(\sigma) = N H \beta (1 - \delta)^2  \mathbb{P}_{\text{imp}}.
\]
In the large $N$ limit,  $N(\sigma'') = N(1-\delta)$ and this  term  gives a contribution to $p_{\text{E}}(\sigma)$ equal to 
\[ 
d \beta^2 (1- \delta) \mathbb{P}_{\text{imp}}.
\]

Let us now look at the second term. Because of the $1 - x_j(\xx')$ term, we are looking at states of the world where agent $j$ has died and is replaced by a new agent with an idiosyncratic probability of the next state $\mathbb{P}'_j(s''=\{1\})$ equal to $z$, which is uniformly distributed between $0$ and $1$, with no memory of the past. Therefore, the sum over $\sigma'$ can be taken independently of the future and gives:
\[
\sum_{\xx'',s''=\{1\}} p(\xx') x_j(\xx'') \mathbb{P}'_j(s''=\{1\}) \sum_{\xx',s'} (1 - x_j(\xx')) Q(\sigma'|\sigma) = \beta \delta (1-\delta) \mathbb{E}[z].
\]
Hence, we find that dying agents give a contribution to $p_{\text{E}}(\sigma)$ equal to 
\[ 
d \beta^2 \delta \frac12,
\] 
where we have replaced $\mathbb{E}[z]$ by $1/2$, and again used the fact that $N(\sigma'') \approx N(1-\delta)$ when $N \gg 1$.

Generalizing to all $\ell \geq 1$ time steps in the future, each agent $j$ can either survive $\ell$ times, with probability $(1-\delta)^\ell$ or die at least once, with probability $1-(1-\delta)^\ell$. In the first case, his/her belief is a martingale. In the second case, the last death cuts all dependence from the past. The calculation above can thus be generalised to give a contribution to $p_{\text{E}}(\sigma)$ equal to:
\[
d \, \beta^\ell \left[(1-\delta)^{\ell-1} \mathbb{P}_{\text{imp}} + 
(1-(1-\delta)^{\ell-1}) \frac12 \right].
\]
Summing over $\ell$  yields our final result for the price of equity in our economy:
\be \label{eq_equitya}
p_{\text{E}} = \frac{d \beta}{2} \left[\frac{2\mathbb{P}_{\text{imp}}-1}{1-\beta(1-\delta)} + \frac{1}{1-\beta} \right].
\ee 
If agents never die, we recover
\[
p_{\text{E}} = d \frac{\beta \mathbb{P}_{\text{imp}}}{1-\beta},
\]
as expected. If agent die at every time step, then $\mathbb{P}_{\text{imp}} \equiv \frac12$ and one also recovers the expected result.

\section{Appendix F: Multiplicative Random Process with Reset}
\label{multiplicative}

Consider the simplest case where, conditioned on survival, returns are IID random variables, i.e.:
 \begin{equation}
 W_i' =  \left\{
 \begin{matrix} 
 W_i (1 + \eta)  & \text{w.p.} & 1-\delta,  \\
       1 & \text{w.p.} & \delta
 \end{matrix} \right.
 \end{equation}
where $\eta$ is the date $t$ element of a sequence of  IID random variables with zero mean and variance equal to $\sigma^2$. For this simple case the sequence of conditional probability measures $\varrho(W)$  obeys the operator equation,
\begin{equation}
 \varrho(W') = (1- \delta) \int {\rm d}W \varrho(W) \int {\rm d}\eta p(\eta) \, {\mathbf{d}}\left(W'-W(1+\eta)\right) + \delta \mathbf{d}(W'-1),   
\end{equation}
where $\mathbf{d}$ is Dirac's delta function. 
For large $W'$ this equation delivers  a power-law tail, with an exponent $\mu$ which is implicitly defined by  the self-consistency condition
\begin{equation} 
 1 = (1 - \delta) \int {\rm d}\eta \, p(\eta) \, (1 + \eta)^\mu.
\end{equation}
In the limit when $\delta$ and $\sigma^2$ are small, the solution for $\mu$ is approximated by the expression,
\be 
\mu = \frac12 \left[ 1 + \sqrt{1+\frac{8\delta}{\sigma^2}} \right].
\ee 
For the wealth process considered in the paper, however, the $\eta$ are correlated in time (since agent $i$ will consistently make/lose money as long as the sign of $\mathbb{P}_i(t)- \mathbb{P}(t)$ is constant, i.e. during a time $\sim \lambda^{-1}$), and its variance is time dependent (see Eqs. \eqref{eq_return} and \eqref{eq_variance}). 

A simplified analysis assumes that $\eta$ is constant during a time $\lambda^{-1}$. This provides the following approximation for $\mu$ in this case:
\begin{equation}\label{eq_power_law2}
\mu \approx \frac12 \left[ 1 + \sqrt{1+\frac{8 \delta \lambda}{\bar{\sigma}^2}} \right],
\qquad {\bar{\sigma}}^2 := \mathbb{E}[\sigma^2(t)].
\end{equation}

Note that $\mu \geq 1$ from this formula, meaning that the wealth distribution always has a finite mean when $\delta >0$. 

A way to decrease wealth inequalities is to introduce a wealth tax. If at each time step a small fraction $\varphi$ of the wealth of each individual is levied and redistributed across the economy, the value of $\mu$ in the simple IID model above changes to:
\[
\mu = \frac{\varphi + \sqrt{\varphi^2 + 2 \delta \sigma^2}}{\sigma^2}.
\]
Hence, as expected, increasing $\varphi$ increases $\mu$ and decreases both the Gini coefficient, thereby making markets more efficient in the sense that the difference between $\mathbb{P}$ and $\mathbb{P}_\text{imp}$ is reduced.

\end{document}